\documentclass[letterpaper, 10 pt, conference]{ieeeconf}  
\IEEEoverridecommandlockouts         % This command is only needed if 
\usepackage{amsmath}  % assumes amsmath package installed
\usepackage{amssymb}  % assumes amsmath package installed
\usepackage{graphicx}
\usepackage{bm}       % % de­fines a com­mand \bm which makes its ar­gu­ment bold 
\usepackage{xcolor}
\newtheorem{theorem}{Theorem}

\newtheorem{remark}{Remark}

\newtheorem{proposition}{Proposition}
\newtheorem{definition}{Definition}

\newcommand*{\qed}{\hfill\ensuremath{\blacksquare}}%
\newcommand{\concatenate}{\mathbin{\rotatebox[origin=c]{90}{$\ominus$}}}
\makeatletter
\newcommand*{\bigconcatenate}{\DOTSB\bigconcatenate@\slimits@}
\newcommand{\bigconcatenate@}{\mathop{\mathpalette\bigconcatenate@@\relax}}
\newcommand{\bigconcatenate@@}[2]{%
  \vcenter{\hbox{%
    \sbox\z@{$\m@th#1\bigoplus$}%
    \resizebox{\wd\z@}{!}{\rotatebox[origin=c]{90}{$\m@th#1\bm{\ominus}$}%
  }}}%
}
\makeatother
\title{\LARGE {\bf
Stabilizability of Markov jump linear systems modeling wireless networked control scenarios
}}
\author{UNIVAQ}\author{Yuriy Zacchia Lun and Alessandro D'Innocenzo % <-this % stops a space
\thanks{Y. Zacchia Lun is with the IMT School for Advanced Studies Lucca, Italy.
A. D'Innocenzo is with the Department of Information Engineering, Computer Science and Mathematics of the University of L'Aquila, Italy.
The research leading to these results has received funding from the 
Sofidel SpA and Tuscany Region under POR FSE 2014-2020, Priority Axis A - Action A.2.1.7 - 
project SINCERA, number 172847, CUP D61J17000000004,
and from the
Italian Government under CIPE resolution n.135 (Dec. 21, 2012), project \emph{INnovating City Planning through Information and Communication Technologies} (INCIPICT).
}% <-this % stops a space
}
\begin{document}

\maketitle
\thispagestyle{empty}
\pagestyle{empty}
%%%%%%%%%%%%%%%%%%%%%%%%%%%%%%%%%%%%%%%%%%%%%%%%%%%%%%%%%%%%%%%%%%%%%%%%%%%%%%%%
%2345678901234567890123456789012345678901234567890123456789012345678901234567890
%        1         2         3         4         5         6         7         8
%%%%%%%%%%%%%%%%%%%%%%%%%%%%%%%%%%%%%%%%%%%%%%%%%%%%%%%%%%%%%%%%%%%%%%%%%%%%%%%%
\begin{abstract}
The communication channels used to convey information between the components of wireless networked control systems (WNCSs) are subject to packet losses due to time-varying fading and interference. The WNCSs with missing packets can be modeled as Markov jump linear systems with one time-step delayed mode observations. While the problem of the optimal linear quadratic regulation for such systems has been already solved, we derive the necessary and sufficient conditions for stabilizability.
We also
%also provide the linear matrix inequality conditions to check the stabilizability, and
show, with an example considering a communication channel model based on WirelessHART (a on-the-market wireless communication standard specifically designed for process automation), that such conditions are essential to the analysis of WNCSs where packet losses are modeled with Bernoulli random variables representing the expected value of the real random process governing the channel.
\end{abstract}
%%%%%%%%%%%%%%%%%%%%%%%%%%%%%%%%%%%%%%%%%%%%%%%%%%%%%%%%%%%%%%%%%%%%%%%%%%%%%%%%
%2345678901234567890123456789012345678901234567890123456789012345678901234567890
%        1         2         3         4         5         6         7         8
%%%%%%%%%%%%%%%%%%%%%%%%%%%%%%%%%%%%%%%%%%%%%%%%%%%%%%%%%%%%%%%%%%%%%%%%%%%%%%%%
\section{INTRODUCTION}
From the automatic control perspective, the wireless communication channels are 
the means to convey information between sensors, actuators, and computational 
units of wireless networked control systems. These communication channels are 
frequently subject to time-varying fading and interference, which may lead to 
packet losses. 

In the wireless networked control system (WNCS) literature the packet dropouts have 
been modeled either as stochastic or deterministic phenomena \cite{Hespanha2007}. 
The proposed deterministic models specify packet losses in terms of time averages 
or in terms of worst case bounds on the number of consecutive dropouts 
(see e.g. \cite{heemels2010networked}). For what concerns stochastic models, 
a vast amount of research assumes memoryless packet drops, so that dropouts are 
realizations of a Bernoulli process (\cite{4118476}, \cite{gupta2009data}, 
\cite{PajicTAC2011}). Other works consider more general correlated (bursty) 
packet losses and use a transition probability matrix (TPM) of a finite-state 
stationary Markov chain (MC, see e.g. the finite-state Markov modelling of 
Rayleigh, Rician and Nakagami fading channels in \cite{sadeghi2008finite} and 
references therein) to describe the stochastic process that rules packet dropouts 
(see \cite{4118476}, \cite{Goncalves20102842}, \cite{baras2008ifac}). 
In these works networked control systems with missing packets are modeled as 
time-homogeneous Markov jump linear systems (MJLSs, \cite{costa2006discrete}), 
with one time-step delayed mode observations. 

While the problem of the optimal linear quadratic regulation for such systems 
has been apparently solved \cite{baras2008ifac}, the existing solution does not consider the aspects of stabilizability. As the main contribution of this article we derive the necessary and sufficient conditions for stabilizability of WNCSs modeled as MJLSs with one time-step delayed mode observations. In addition to this, we provide an example considering a communication channel model based on WirelessHART (a on-the-market wireless communication standard specifically designed for process automation) and show that the stationary MJLS model derived from the accurate Markov channel representation of the communication channel permits to discover and overcome the challenging subtleties arising from bursty behavior. We also show that our stabilizability conditions are essential to the analysis of WNCSs that consider Bernoulli dropouts, when the Bernoulli random variables represent the expected value of the real random process governing the packet losses.

%%%%%%%%%%%%%%%%%%%%%%%%%%%%%%%%%%%%%%%%%%%%%%%%%%%%%%%%%%%%%%%%%%%%%%%%%%%%%%%%
\textbf{Notation and preliminaries}. In the following, 
{\small $\mathbb{N}_{0}$} denotes the set of non-negative integers, while 
{\small $\mathbb{F}$} indicates the set of either real or complex numbers.
The absolute value of a number is denoted by {\small $|\cdot|$}.
%%%%%%%%%%%%%%%%%%%%%%%%%%%%%%%%%%%%%%%%%%%%%%%%%%%%%%%%%%%%%%%%%%%%%%%%%%%%%%%%
We recall that every finite-dimensional normed space over {\small $\mathbb{F}$} 
is 
% a complete normed vector space, i.e.
a Banach space \cite{megginson1998introduction}, 
%%%%%%%%%%%%%%%%%%%%%%%%%%%%%%%%%%%%%%%%%%%%%%%%%%%%%%%%%%%%%%%%%%%%%%%%%%%%%%%%
and denote the Banach space of all bounded linear operators of Banach space 
{\small $\mathbb{X}$} into Banach space {\small $\mathbb{Y}$}, by 
{\small $\mathbb{B}\!\left(\mathbb{X},\mathbb{Y}\right)$}, and set
{\small $\mathbb{B}\!\left(\mathbb{X},\mathbb{X}\right)\!\triangleq\!
\mathbb{B}\!\left(\mathbb{X}\right)$}.
%%%%%%%%%%%%%%%%%%%%%%%%%%%%%%%%%%%%%%%%%%%%%%%%%%%%%%%%%%%%%%%%%%%%%%%%%%%%%%%%
The identity matrix of size {\small $n$} is indicated by {\small 
$\mathbb{I}_{n}$}. The operation of transposition is denoted by apostrophe, 
the complex conjugation by overbar, while the conjugate transposition is indicated 
by superscript {\small $^*$}, the real part of the elements of the complex matrix
by {\small $\Re\!\left(\cdot\right)$}. % Clearly, 
% for a set of real matrices, the transpose and conjugate transpose are the same.
%%%%%%%%%%%%%%%%%%%%%%%%%%%%%%%%%%%%%%%%%%%%%%%%%%%%%%%%%%%%%%%%%%%%%%%%%%%%%%%%
We denote by {\small $\rho(\cdot)$} the spectral radius of a square matrix 
(or a bounded linear operator), i.e., the largest absolute value of its eigenvalues, 
and by {\small 
$\left\| \cdot \right\|$} either any vector norm or any matrix norm.
Since for finite-dimensional linear spaces all norms are equivalent
\cite[Theorem 4.27]{kubrusly2001elements} from a topological viewpoint, as vector 
%%%%%%%%%%%%%%%%%%%%%%%%%%%%%%%%%%%%%%%%%%%%%%%%%%%%%%%%%%%%%%%%%%%%%%%%%%%%%%%%
norms we use variants of vector $p$-norms.
For what concerns the matrix norms, we use % two kinds of entry-wise norms,
{\small $\ell_1$} and {\small $\ell_2$} norms \cite[p.~341]{horn2012matrix},
that treat {\small $n_r\!\times\! n_c$} matrices as vectors of size
{\small $n_r n_c$}, and use one of the related $p$-norms. The definition of 
{\small $\ell_1$} and {\small $\ell_2$} norms is based on the operation of 
vectorization of a matrix, {\small $\mathrm{vec}(\cdot)$}, which is further used 
in the definition of the operator {\small $\mathrm{vec}^{2}(\cdot)$}, to be applied 
to any block matrix, e.g.~{\small ${\bm \Phi}\!=\!\big[\Phi_{_{\!\!\;ij}}\big]_{i,j=1}^{N}$}, 
as if its blocks {\small $\Phi_{_{\!\!\;ij}}$} of size {\small $n_r\!\times\! n_c$} 
were the simple elements: 

\vspace*{-4mm}
\begin{small}
\begin{equation*}
\mathrm{vec}^{2}\!\left({\bf\Phi}\right)\!\!\,\triangleq\!\!\,
\left[\mathrm{vec}\!\left({\Phi}_{_{\!\!\;11}}\!\!\;\right)\!,\dots,\mathrm{vec}\!\left({\Phi}_{_{\!\!\;N1}}\!\!\;\right)\!,\mathrm{vec}\!\left({\Phi}_{_{\!\!\;12}}\!\!\;\right)\!,\dots,\mathrm{vec}\!\left({\Phi}_{_{\!\!\;NN}}\!\!\;\right)\right]'\!.
\end{equation*}
\end{small}

\vspace*{-5mm}
\noindent
%%%%%%%%%%%%%%%%%%%%%%%%%%%%%%%%%%%%%%%%%%%%%%%%%%%%%%%%%%%%%%%%%%%%%%%%%%%%%%%%
The linear operator {\small $\mathrm{vec}^{2}(\cdot)$} is a uniform homeomorphisms, 
% % i.e., it is invertible, 
its inverse operator {\small $\mathrm{vec}^{-2}(\cdot)$} 
is uniformly continuous \cite{naylor2000linear},
% %  for additional details),
%%%%%%%%%%%%%%%%%%%%%%%%%%%%%%%%%%%%%%%%%%%%%%%%%%%%%%%%%%%%%%%%%%%%%%%%%%%%%%%%
and any bounded linear operator %%%{\small $\mathcal{L}\!\left(\cdot\right)$} 
in {\small $\mathbb{B}\!\left(\mathbb{F}^{Nn_r\times Nn_c}\right)$} can be 
represented in {\small $\mathbb{B}\!\left(\mathbb{F}^{N^{2}n_{r}n_c}\right)$} trough 
{\small $\mathrm{vec}^{2}(\cdot)$}.
%Thus, as an immediate adaptation of \cite[Lemma 1, p. 1085]{} in
%%%%%%%%%%%%%%%%%%%%%%%%%%%%%%%%%%%%%%%%%%%%%%%%%%%%%%%%%%%%%%%%%%%%%%%%%%%%%%%%
Then, $\concatenate$ indicates the operation of matrix augmentation, i.e., 
the horizontal concatenation of two matrices with the same number of rows. 
%%%%%%%%%%%%%%%%%%%%%%%%%%%%%%%%%%%%%%%%%%%%%%%%%%%%%%%%%%%%%%%%%%%%%%%%%%%%%%%%%%
We denote by {\small $\otimes$} the Kronecker product defined in the 
usual way, see e.g. \cite{brewer1978kronecker}, and by $\oplus$ the direct sum. 
Notably, the direct sum of a sequence of square matrices 
{\small $\left( \Phi_i \right)_{i = 1}^{N}$} produces a block diagonal matrix, 
having its elements,  {\small $\Phi_i$}, on the main diagonal blocks. 
Then, {\small $\mathrm{trace}\left(\cdot\right)$} indicates the trace of a square 
matrix. For two Hermitian matrices of the same dimensions, 
{\small $\Phi_{_{\!1}}$} and {\small $\Phi_{_{\!2}}$}, 
{\small $\Phi_{_{\!1}}\!\succeq\!\Phi_{_{\!2}}$} (respectively 
{\small $\Phi_{_{\!1}}\!\succ\!\Phi_{_{\!2}}$}) means that 
{\small $\Phi_{_{\!1}}\!-\!\Phi_{_{\!2}}$} is positive semi-definite
(respectively positive definite). 
Finally, {\small $\mathbb{E}\!\left(\cdot\right)$} stands for the mathematical 
expectation of the underlying scalar valued random variables.

%%%%%%%%%%%%%%%%%%%%%%%%%%%%%%%%%%%%%%%%%%%%%%%%%%%%%%%%%%%%%%%%%%%%%%%%%%%%%%%%
\section{COMMUNICATION CHANNEL MODEL}\label{sec:whart}
%%%%%%%%%%%%%%%%%%%%%%%%%%%%%%%%%%%%%%%%%%%%%%%%%%%%%%%%%%%%%%%%%%%%%%%%%%%%%%%%
%2345678901234567890123456789012345678901234567890123456789012345678901234567890
%        1         2         3         4         5         6         7         8
%%%%%%%%%%%%%%%%%%%%%%%%%%%%%%%%%%%%%%%%%%%%%%%%%%%%%%%%%%%%%%%%%%%%%%%%%%%%%%%%
The challenges in analysis and co-design of WNCSs are best explained by considering 
wireless industrial control protocols. In this paper we focus on a networking protocol 
specifically developed for wireless industrial automation, i.e. WirelessHART 
\cite{IEC62591:2016}, which is based upon the physical layer of IEEE 802.15.4-2006.
Since interleaving and forward error correction techniques appear only in the 
IEEE 802.15.4-2015 version of the standard, even one erroneous bit leads to a corrupted 
WirelessHART data packet. According to the IEEE standard \cite[p.~268]{2006ieee}, the bit 
error ratio (BER, {\small $\mathrm{R}_{\mathrm{b}}$}) depends only on the 
signal-to-noise-plus-interference ratio (SNIR),
that may be expressed by the sum of weighted log-normal processes. Such model admits
an accurate log-normal approximation based on moment matching method \cite{fischione2007}.
In the following we will indicate this approximation as analytic model.
Notably, using a logarithmic scale for the values of SNIR, denoted by {\small $\Gamma$}, 
gives a normally distributed probability density function with mean {\small $\mu$} and variance {\small $\sigma^2$}, 
{\small $\Gamma \sim\mathcal{N}\!\left(\mu,\sigma^2\right)$}. 
For notational convenience, we will indicate by {\small $\gamma$} the power value 
corresponding to SNIR {\small $\Gamma$}, i.e.~{\small $\Gamma\!\triangleq\!10\log_{10}\left(\gamma\right)$} [dB].

%%%%%%%%%%%%%%%%%%%%%%%%%%%%%%%%%%%%%%%%%%%%%%%%%%%%%%%%%%%%%%%%%%%%%%%%%%%%%%%%
An analytic model may be used to create a finite-state Markov channel model \cite{sadeghi2008finite}
that captures the essence of time-varying channel behavior by associating a binary symmetric channel to
each state of an ergodic discrete-time MC. Clearly, the approximation becomes more 
accurate as the number of MC states becomes larger. Conversely, the coarsest approximation 
of the channel behavior looks at only one state and may completely neglect second-order statistics.
It considers the packet error probability % corresponding to \textcolor{red}{expected value of} the SNIR 
to follow a Bernoulli distribution. This simple model has been widely adopted in the WNCSs literature.
% (\cite{4118476}, \cite{gupta2009data}, \cite{PajicTAC2011}). 
In Section~\ref{sec:stabilizability} we will show that when the derived 
(strong) stabilizability conditions are satisfied, the linear quadratic regulation with 
Bernoulli dropouts \cite{4118476} provides a mode-independent solution to the optimal infinite-horizon 
state feedback control problem over a more rigorous representation of the wireless communication channel.
Here, we show a link between the stochastic and deterministic models by
deriving worst case bounds on the number of consecutive dropouts.

%%%%%%%%%%%%%%%%%%%%%%%%%%%%%%%%%%%%%%%%%%%%%%%%%%%%%%%%%%%%%%%%%%%%%%%%%%%%%%%%
% \subsection{Analytic model} 
%%%%%%%%%%%%%%%%%%%%%%%%%%%%%%%%%%%%%%%%%%%%%%%%%%%%%%%%%%%%%%%%%%%%%%%%%%%%%%%%
\textbf{Analytic model.} The reference scenario is given by a certain number of coexisting and independent
WirelessHART networks. For simplicity, here we assume that there are only two networks,
and the transmitted signals are affected by path loss \cite[p.~274]{2006ieee}, 
(log-normal) shadowing, and additive white noise introduced by the channel. 
Since in industrial setting highly absorbing environments eliminate multipath propagation 
\cite{barac2014scrutinizing}, the multipath-induced fading is neglected. The considered modulation 
scheme is offset quadrature phase-shift keying direct-sequence spread spectrum, and it 
is supported by a coherent demodulation. To convey the control system data,
WirelessHART uses Publish data messages \cite[p.~248]{IEC62591:2016}, where
the minimum update period is $0.1$\,s, time slot period {\small $\mathrm{T}_{\mathrm{s}}$} is
$0.01$\,s, and the frame length {\small $\mathrm{L}_{\mathrm{F}}$} for e.g.~four relevant 
state variables is $26$\,octets. % , i.e.~{\small $208$}\,bits.
From \cite[p.~268]{2006ieee} and the absence of the forward error correction, the 
packet error rate (PER, {\small $\mathrm{R}_{\mathrm{p}}$}) is related to SNIR through BER,
where {\small $\gamma\!\in\!\left(0,\infty\right)$}, and
{\small $\mathrm{R}_{\mathrm{b}}\!\left(\gamma\right)\!\in\!\left[0,0.5\right]$},
{\small $\mathrm{R}_{\mathrm{p}}\!\left(\gamma\right)\!\in\!\left[0,1\right]$}
are both monotonically non-increasing:

\vspace*{-9.72mm}
%%%%%%%%%%%%%%%%%%%%%%%%%%%%%%%%%%%%%%%%%%%%%%%%%%%%%%%%%%%%%%%%%%%%%%%%%%%%%%%%
\begin{small}
\begin{align*}\label{eq:PER}
\hspace*{2cm}\mathrm{R}_{\mathrm{p}}\!\left(\gamma\right)&=\!1\!-\!\left(1\!-\!\mathrm{R}_{\mathrm{b}}\!\left(\gamma\right)\right)^{\mathrm{L}_{\mathrm{F}}}\!,~\text{where}\\
\mathrm{R}_{\mathrm{b}}\!\left(\gamma\right)&=\!\frac{1}{30}\!\sum_{i=2}^{16}\!\left(-1\right)^{i}\!\!\binom{16}{i}e^{\left(20 \gamma \frac{1-i}{i} \right)}.\nonumber
\end{align*}
\end{small}

\vspace*{-3mm}
\noindent 
So, for {\small $\mathrm{L}_{\mathrm{F}}\!=\!208$}, we have that
{\small $\mathrm{R}_{\mathrm{p}}\!<\!2.22\cdot10^{-16}$} 
{\small $\forall \gamma^{\diamond}\!\geq\!3.882$} (that is, {\small $\Gamma^{\diamond}\!\geq\!5.89$} dB), and 
{\small $\mathrm{R}_{\mathrm{p}}\!<\!3.17\cdot10^{-10}$} (i.e, a rate of less than 
1 data packet lost in a year of continuous operation with the sampling time {\small $\mathrm{T}_{\mathrm{s}}$}) 
{\small $\forall \gamma^{\star}\!\geq\!2.859$} (i.e., {\small $\Gamma^{\star}\!\geq\!4.56$} dB).

Since in practical applications the probability of packet error 
burst of length {\small $\mathrm{L}_{\mathrm{B}}$} is negligible when
it is smaller than a specified threshold {\small $\varepsilon$},
which may be as small as the \textit{machine epsilon}, the highest number of consecutive dropouts {\small $\mathrm{L}_{\mathrm{B}}\!\left(\varepsilon\right)$} may be obtained as follows.
As {\small $\Gamma \sim\mathcal{N}\!\left(\mu,\sigma^2\right)$}, its 
cumulative distribution function {\small $F_{_{\!\Gamma}}\!\left(\varepsilon\right)$}
gives the probability of a single packet loss, while 
{\small $F_{_{\!\Gamma}}\!\left(\varepsilon\right)^{\mathrm{L}_{\mathrm{B}}}$} is the
probability of {\small $\mathrm{L}_{\mathrm{B}}$} consecutive packet losses, so 

\vspace*{-2mm}
%%%%%%%%%%%%%%%%%%%%%%%%%%%%%%%%%%%%%%%%%%%%%%%%%%%%%%%%%%%%%%%%%%%%%%%%%%%%%%%%
\begin{small}
\begin{equation}\label{eq:LB}
\mathrm{L}_{\mathrm{B}}\!\left(\varepsilon\right)\!=\!
	\mathrm{ceil}\!\left(\frac{\ln\!\left(\varepsilon\right)}
		{\ln\!\left(\frac{1}{2}\!\left(1\!+\!\mathrm{erf}\!\left(
		\frac{\Gamma\left(\varepsilon\right)-\mu}{\sigma\sqrt{2}}\right)\!\right)
		\!\right)}\!\right)\!,
\end{equation}
\end{small}
\vspace*{-3mm}

\noindent where {\small $\mathrm{ceil}\!\left(\cdot\right)$}, 
{\small $\ln\!\left(\cdot\right)$} and {\small $\mathrm{erf}\!\left(\cdot\right)$}
are the ceiling, natural logarithm, and error functions, respectively. 

As an illustrative example, consider {\small $\varepsilon^{\star}\!=\!3.17\cdot10^{-10}$}, so that 
{\small $\Gamma\!\left(\varepsilon^{\star}\right)\!=\!\Gamma^{\star}\!=\!4.56$}~dB, on a
wireless communication channel with {\small $\hat{\mu}\!=10.15\!$}~dB and {\small $\hat{\sigma}\!=4.85\!$}~dB.
Then, {\small $\mathrm{L}_{\mathrm{B}}\!\left(\varepsilon^{\star}\right)\!=\!11$}. For 
{\small $\varepsilon^{\diamond}\!=\!2.22\cdot10^{-16}$} and the same channel characterized by
{\small $\hat{\mu}$}, {\small $\hat{\sigma}$}, we have instead that 
{\small $\Gamma\!\left(\varepsilon^{\diamond}\right)\!=\!\Gamma^{\diamond}\!=\!5.89\!$}~dB, so that
{\small $\mathrm{L}_{\mathrm{B}}\!\left(\varepsilon^{\diamond}\right)\!=\!22$}.

Since the probability density function {\small $f_{\Gamma}\!\left(\cdot\right)$} of the model is known,
and {\small $\mathrm{R}_{\mathrm{p}}$} is a continuous function defined on the range of {\small $\Gamma$},
by the law of the unconscious statistician, the expected value of the PER can be obtained as
{\small $\int_{-\infty}^{+\infty} 
\mathrm{R}_{\mathrm{p}}(10^{\frac{\alpha}{10}}) f_{\Gamma}\!\left(\alpha\right)d\alpha$},
and its variance can be derived in a similar fashion. On the channel 
{\small $\hat{\Gamma} \sim\mathcal{N}\!\left(\hat{\mu},\hat{\sigma}^2\right)$} having
{\small $\hat{\mu}\!=10.15\!$}~dB and {\small $\hat{\sigma}\!=4.85\!$}~dB,
{\small $\mathbb{E}(\mathrm{R}_{\mathrm{p}}(\hat{\Gamma}))\!=\!0.008$},
while the PER variance equals to {\small $0.006$}.

%\subsection{Finite-state abstractions}
\textbf{Finite-state abstractions.}
The analytic model of a channel % % with SNIR {\small $\Gamma \sim\mathcal{N}\!\left(\mu,\sigma^2\right)$}
is defined on continuous state-space. However, there are several application scenarios 
(e.g.~modeling channel error bursts, decoding in channels with memory, adaptive transmission) 
where using a finite number of channel states can be more advantageous \cite{sadeghi2008finite}.
%%%
The coarsest abstraction of the analytic model collapses the infinite-dimensional state-space into one state
with a representative PER (given by its expected value), which may be seen as a probability of the 
packet loss event in the Bernoulli distribution.
%%% 
In more accurate finite-state Markov channel abstraction the range of SNIR 
is divided into several consecutive regions. A region {\small $i$} is mapped
into a state {\small $s_i$} of the related MC and is delimited by two thresholds
{\small $\alpha_i$} and {\small $\alpha_{i+1}$}. The steady state probability of
the state {\small $s_i$} is the probability that 
the SNIR is between the thresholds of the region, 
i.e.~{\small $\mathbf{p}_{i}\!=\!\int_{\alpha_i}^{\alpha_{i+1}}\!f_{\Gamma}(\alpha)d\alpha$},
while the PER associated to the same state is given by its expected value 
within the respective region, that is
{\small $\mathrm{R}_{\mathrm{p}}^{i}\!=\!\frac{1}{\mathbf{p}_{i}}\!\int_{\alpha_i}^{\alpha_{i+1}}\!
\mathrm{R}_{\mathrm{p}}(10^{\frac{\alpha}{10}}) f_{\Gamma}\!\left(\alpha\right)d\alpha$}.
The TPM may be then obtained by integrating the joint probability density function of the SNIR 
\cite{sadeghi2008finite} over two consecutive packet transmissions and over the desired regions.
In the literature on finite-state Markov channel abstractions there are different 
methods of partitioning of the range of SNIR \cite{sadeghi2008finite,ruiz2009finite}.
For simplicity of the presentation, in this paper we choose {\small $\Gamma(\varepsilon)$} 
as the only threshold. {\small $\Gamma(\varepsilon)$} divides the range of SNIR in two intervals. 
In this way we obtain a Markov channel with just two operational modes, where only one mode of operation
has nonzero packet error probability. This model is known as Gilbert channel \cite{sadeghi2008finite}.
It is the easiest nontrivial example of channel models with memory.

%%%%%%%%%%%%%%%%%%%%%%%%%%%%%%%%%%%%%%%%%%%%%%%%%%%%%%%%%%%%%%%%%%%%%%%%%%%%%%%%
\section{OPTIMAL CONTROL SCHEMES}
%%%%%%%%%%%%%%%%%%%%%%%%%%%%%%%%%%%%%%%%%%%%%%%%%%%%%%%%%%%%%%%%%%%%%%%%%%%%%%%%
%2345678901234567890123456789012345678901234567890123456789012345678901234567890
%        1         2         3         4         5         6         7         8
%%%%%%%%%%%%%%%%%%%%%%%%%%%%%%%%%%%%%%%%%%%%%%%%%%%%%%%%%%%%%%%%%%%%%%%%%%%%%%%%
Consider a linear stochastic system with intermittent control 
packets due to the lossy communication channel \cite{4118476}

\vspace*{-2mm}
%%%%%%%%%%%%%%%%%%%%%%%%%%%%%%%%%%%%%%%%%%%%%%%%%%%%%%%%%%%%%%%%%%%%%%%%%%%%%%%%
\begin{small}
\begin{equation}\label{eq:control}
x_{k+1}\!=\!A x_k \!+\! B u_k^a \!+\! w_k,\quad\text{with}\quad
u_k^a\!=\!\nu_k u_k^c,
\end{equation}
\end{small}
%%%%%%%%%%%%%%%%%%%%%%%%%%%%%%%%%%%%%%%%%%%%%%%%%%%%%%%%%%%%%%%%%%%%%%%%%%%%%%%%

\vspace*{-5mm}
\noindent
where, {\small $x_k\!\in\!\mathbb{F}^{n_x}$} is a system state, 
{\small $u_k^a\!\in\!\mathbb{F}^{n_u}$} is the control input to the actuator, 
{\small $A$} and {\small $B$} are state and input matrices of appropriate size, 
respectively, {\small $u_k^c\!\in\!\mathbb{F}^{n_u}$} is the desired control input 
computed by the controller, {\small $w_k\!\in\!\mathbb{F}^{n_x}$} is a Gaussian white 
process noise with zero mean and covariance 
matrix {\small $\Sigma_w$}. The process noise {\small $w_k$} is assumed to be
independent from the initial state $x_0$ and from the stochastic variable $\nu_k$, 
which models the packet loss between the controller and the actuator: if the packet 
is correctly delivered then {\small $u_k^a\!=\!u_k^c$}, otherwise if it is lost then the 
actuator does nothing, i.e., {\small $u_k^a\!=\!0$}. This compensation scheme is 
summarized by \eqref{eq:control}. 

In the following, we assume full state observation with no measurement noise, 
and no observation packet loss, so the optimal control must 
necessarily be a static state feedback and no filter is necessary. 
In such setting, we will compare the performance of 
the optimal state feedback controller under TCP-like protocols \cite{4118476} 
(treating {\small $\nu_k$} as independent and 
identically distributed (i.i.d.) Bernoulli random variables, with information 
set available to the controller defined as
{\small $\mathcal{F}_k\!\triangleq\!\left\{{\bm x}^k, {\bm \nu}^{k-1} \right\}$}, 
where {\small ${\bm x}^k\!=\!\left(x_t\right)_{t=0}^k$}, and 
{\small ${\bm \nu}^{k}\!=\!\left(\nu_{t}\right)_{t=0}^k$}, and
optimal linear quadratic regulator for MJLS, in the presence of one time-step 
delayed mode observations \cite{baras2008ifac}, which considers {\small $\nu_k$} 
as a random variable governed by the Markov channel derived from the model of Section \ref{sec:whart}.

%%%%%%%%%%%%%%%%%%%%%%%%%%%%%%%%%%%%%%%%%%%%%%%%%%%%%%%%%%%%%%%%%%%%%%%%%%%%%%%%
\textbf{Linear quadratic regulation with Bernoulli dropouts.} 
The optimal state-feedback controller accounting for the i.i.d. packet losses following a Bernoulli distribution,  with 
{\small $\mathrm{Pr}\!\left(\nu_k\!=\!1\right)\!=\!\hat{\nu}$} for all {\small $k$}, 
minimizes the performance index {\small $J^b\!=\!\lim_{t\to\infty}\!
	\frac{1}{t}\mathbb{E}(
	\sum_{k=0}^{t}(x_{k}^{*} Q x_{k}\!+\!u_{k}^{b*}\!Ru_{k}^b
	)\mid \mathcal{F}_k)$},
%%%%%%%%%%%%%%%%%%%%%%%%%%%%%%%%%%%%%%%%%%%%%%%%%%%%%%%%%%%%%%%%%%%%%%%%%%%%%%%%
%\begin{small}
%\begin{equation*}
%J^b\!=\!\lim_{t\to\infty}\!
%	\frac{1}{t}\mathbb{E}\!\left(
%	\sum\limits_{k=0}^{t}\left(x_{k}^{*} Q x_{k}\!+\!u_{k}^{b*}\!Ru_{k}^b
%	\right)\mid \mathcal{F}_k\right).
%\end{equation*}
%\end{small}
%%%%%%%%%%%%%%%%%%%%%%%%%%%%%%%%%%%%%%%%%%%%%%%%%%%%%%%%%%%%%%%%%%%%%%%%%%%%%%%%
where {\small $Q\!\succeq\!0$} and {\small $R\!\succ\!0$} are the state
weighting and control weighting matrices, respectively. 
The optimal controller is given by {\small $u_k^b\!=
	\!-(R\!+\!B^{*} X_{\infty}^b B)^{-1} (B^{*} X_{\infty}^b A) x_k \!=\! K^b x_k$},
%%%%%%%%%%%%%%%%%%%%%%%%%%%%%%%%%%%%%%%%%%%%%%%%%%%%%%%%%%%%%%%%%%%%%%%%%%%%%%%%
%\begin{small}
%\begin{equation}\label{eq:modifiedLQR}
%u_k^b\!=
%	\!-(R\!+\!B^{*} X_{\infty}^b B)^{-1} (B^{*} X_{\infty}^b A) x_k \!=\! K^b x_k,
%\end{equation}
%\end{small}
%%%%%%%%%%%%%%%%%%%%%%%%%%%%%%%%%%%%%%%%%%%%%%%%%%%%%%%%%%%%%%%%%%%%%%%%%%%%%%%%
where {\small $X_{\infty}^b$} is the unique positive semi-definite solution
of the modified algebraic Riccati equation (MARE, \cite{4118476})
{\small $X_{\infty}^b\!=\!A^{*}X_{\infty}^b A\!-\!
	\hat{\nu}(A^{*}\! X_{\infty}^b B)(R\!+\!B^{*}\!
	X_{\infty}^b B)^{{-1}}\!(B^{*}\! X_{\infty}^b A)\!+\!Q$}.
%\begin{small}
%\begin{equation*}
%X_{\infty}^b\!=\!A^{*}X_{\infty}^b A\!-\!\hat{\nu}(A^{*} X_{\infty}^b B)(R\!+\!B^{*} 
%	X_{\infty}^b B)^{{-1}}(B^{*} X_{\infty}^b A)\!+\!Q.
%\end{equation*}
%\end{small}
%%%%%%%%%%%%%%%%%%%%%%%%%%%%%%%%%%%%%%%%%%%%%%%%%%%%%%%%%%%%%%%%%%%%%%%%%%%%%%%%
%%%For $Q\!\succeq\!0$ and $R\!\succ\!0$, i
If{\small $\;\left(A,B\right)$} is controllable, 
and {\small $\left(A,Q\right)$} is observable, the solution {\small $X_{\infty}^b$} 
is stabilizing if and only if {\small $\hat{\nu}\!>\!\nu_c$} (see \cite{4118476} 
together with \cite{7109127}), where {\small $\nu_c$} is the critical arrival probability, that 
satisfies the following analytical bound: {\small $\nu_c \!\leq\! p_{\max}$}, with {\small $p_{\max}\!\triangleq\!1\!-\!{\Pi_{i}|\lambda_i^u(A)|^{-2}}$},
%%%%%%%%%%%%%%%%%%%%%%%%%%%%%%%%%%%%%%%%%%%%%%%%%%%%%%%%%%%%%%%%%%%%%%%%%%%%%%%%
%\begin{small}
%\begin{equation*}
%p_{\max}\!\triangleq\!1\!-\!\frac{1}{\Pi_{i}|\lambda_i^u(A)|^{2}},
%\end{equation*}
%\end{small}
%%%%%%%%%%%%%%%%%%%%%%%%%%%%%%%%%%%%%%%%%%%%%%%%%%%%%%%%%%%%%%%%%%%%%%%%%%%%%%%%
where {\small $\lambda_i^u(A)$} are unstable eigenvalues of {\small $A$}. 
Notably, when {\small $B$} is rank {\small $1$}, {\small $\nu_c \!=\! p_{\max}$}.
%%%%%%%%%%%%%%%%%%%%%%%%%%%%%%%%%%%%%%%%%%%%%%%%%%%%%%%%%%%%%%%%%%%%%%%%%%%%%%%%
The optimal controller attains the optimal value of the performance 
index, i.e., %%%which is given by
{\small $J^b_{\star}\!=\!\mathrm{trace}\!\left(X_{\infty}^b\Sigma_w\right)$}.

%%%%%%%%%%%%%%%%%%%%%%%%%%%%%%%%%%%%%%%%%%%%%%%%%%%%%%%%%%%%%%%%%%%%%%%%%%%%%%%%
\textbf{MJLS with one time-step delayed mode observations.} 
When the finite-state Markov channel representation of the communication channel 
is available, to each state {\small $s_i$} of the MC (characterising the 
evolution of the channel) is associated a discrete memoryless channel. 
The networked control system using such communication channel can be modelled as
Markov jump system in the following manner.  

%%%%%%%%%%%%%%%%%%%%%%%%%%%%%%%%%%%%%%%%%%%%%%%%%%%%%%%%%%%%%%%%%%%%%%%%%%%%%%%%
Consider the stochastic basis 
{\small $\left(\Omega,\mathcal{G},\left(\mathcal{G}_k\!\right)\!,\mathrm{Pr}\right)$},
where 
{\small $\Omega$} is the sample space, 
{\small $\mathcal{G}$} is the $\sigma$-algebra of (Borel) measurable events,
{\small $\left( \mathcal{G}_k \right)$} is the related filtration, and 
$\mathrm{Pr}$ is the probability measure. Then, the communication channel state 
is the output of the discrete-time MC 
{\small $\Theta \!:\! \mathbb{N}_0 \!\times\! \Omega \!\to\! \mathbb{S}$} defined on the 
probability space, that takes values in a finite set 
{\small $\mathbb{S} \!\triangleq\! \{ s_i \}_{i=1}^N$}. 
{\small $\forall k \!\in\! \mathbb{N}_0$} the transition probability between 
the channel's states is defined as {\small $p_{ij} \!=\! \mathrm{Pr}\{\theta_{k+1} \!=\! s_j \mid \theta_{k} \!=\! s_i \} 
   \geq 0$, $\sum_{j=1}^N p_{ij} \!=\! 1$}.
The associated TPM is a stochastic 
{\small $N \times N$} matrix with entries {\small $p_{ij}$}, i.e., 
{\small $\begin{bmatrix} p_{ij} \end{bmatrix}_{i,j=1}^{N}$}. 
The probability of the successful packet delivery and packet loss are 
now conditioned to the state of the communication 
channel, i.e. respectively

\vspace*{-3.2mm}
\begin{small}
\begin{equation}\label{eq:nu_i}
\mathrm{Pr}\!\left(\nu_k\!=\!1\mid\theta_k\!=\!s_i\right)\!=\!\hat{\nu}_i,\quad
\mathrm{Pr}\!\left(\nu_k\!=\!0\mid\theta_k\!=\!s_i\right)\!=\!1\!-\!\hat{\nu}_i. 
\end{equation}
\end{small}
\vspace*{-5mm}

%%%%%%%%%%%%%%%%%%%%%%%%%%%%%%%%%%%%%%%%%%%%%%%%%%%%%%%%%%%%%%%%%%%%%%%%%%%%%%%%
We observe that the networks based on IEEE 802.15.4 compatible hardware provide
a SNIR estimation procedure, which is performed during link quality indicator 
measurement \cite[p.~65]{2006ieee}. The estimated value of the SNIR falls into 
one of the regions in which the range of SNIR is partitioned.
In this way the corresponding state {\small $s_i$} of the Markov channel is 
unequivocally determined, and {\small $\hat{\nu}_i \!=\! 1\!-\!\mathrm{R}_{\mathrm{p}}^{i}$}.

We present the systems in terms of the Markov framework \cite{costa2006discrete} via the 
augmented state {\small $\left(x_k,\left(\theta_k,\nu_k\right)\right)$}, where 
{\small $\nu_k$} accounts for two possible operational modes, i.e., in closed loop, 
when {\small $\nu_k\!=\!1$}, and in open loop, if {\small $\nu_k\!=\!0$}, while 
{\small $\theta_k$} describes the channel evolution. Since the probability of a 
particular operational mode depends on the state of the communication channel, 
we denote the aggregated state {\small $\left(\theta_k,\nu_k\right)$} by 
{\small $\nu_{\theta_k}$}, which is a {\small $2N$}-ary random quantity. 

The operational modes are observed by controller via acknowledgements, 
which are available only after the current decision on the gain to apply has 
been made and sent through the channel, because the actual success of the 
transmission is not known in advance. We assume that the acknowledgments, and 
also the communication channel states (measured through SNIR), are not received 
at the controller instantaneously, but become available before the next decision 
on the control to apply. So, we are dealing with one time-step delayed mode 
observations, as presented in \cite{baras2008ifac}, and the informational set 
available to the controller is 
{\small $\mathcal{G}_k\!=\!\left\{{\bm x}^k, {\bm \nu}_{\bm \theta}^{k-1}\right\}$},
with {\small ${\bm x}^k\!=\!\left(x_t\right)_{t=0}^k$}, and 
{\small ${\bm \nu}_{\bm \theta}^{k}\!=\!\left(\theta_t,\nu_t\right)_{t=0}^{k}$}.
The state space representation of the system is %% then written as
% First, we rewrite \eqref{eq:state} and \eqref{eq:control} as

\vspace*{-2.3mm}
\begin{small}
\begin{equation}\label{eq:mjls}
x_{k+1} \!=\! A x_k \!+\! \nu_{\theta_k} B u_k \!+\! w_k,
\end{equation}
\end{small}
\vspace*{-5.5mm}

\noindent where the Gaussian process noise {\small $w_k$} (having the zero mean and
covariance matrix {\small $\Sigma_w$}) is assumed to be independent from
the initial condition {\small $\left(x_0,\nu_{\theta_0}\right)$} and from 
the Markov process {\small $\nu_{\theta_k}$} for all values of 
discrete time {\small $k$}.
%%%%%%%%%%%%%%%%%%%%%%%%%%%%%%%%%%%%%%%%%%%%%%%%%%%%%%%%%%%%%%%%%%%%%%%%%%%%%%%%
We make the assumption that the MC {\small $\Theta$} is ergodic, with the 
steady state distribution $\mathbf{p}_{i}$ defined as 
{\small $\lim_{k\to\infty} \mathrm{Pr}\!\left(\theta_k\!=\!s_i\right)$}, 
so the aggregated Markov process {\small $\nu_{\theta_k}$} is also ergodic. 

The optimal mode-dependent infinite-horizon state feedback controller is obtained 
from the solution of the following coupled algebraic Riccati equations (CAREs), 
that are constructed via the general procedure described in \cite{baras2008ifac}, 
after few manipulations that take into account the particular structure 
\eqref{eq:mjls} of the considered MJLS:

\vspace*{-3mm}
\begin{small}
\begin{equation}\label{eq:care}
X_{\infty,i}^c\!=\!\mathcal{A}_i \!-\! \mathcal{C}_i \mathcal{B}_i^{-1} \mathcal{C}_{i}^{*}, 
\quad X_{\infty,i}^c \!=\! X_{\infty,i}^{c*}\!\succeq\!0,
\end{equation}

\vspace*{-6mm}
\begin{equation*}
\mathcal{A}_i\!=\!A^{*}\!\!\left(\sum\nolimits_{j=1}^{N}\!p_{ij}X_{\infty,j}^{c}\!\right)\!A + Q,\quad
\mathcal{C}_i\!=\!A^{*}\!\!\left(\sum\nolimits_{j=1}^{N}\!p_{ij}\hat{\nu}_{j}X_{\infty,j}^{c}\!\right)\!\!B,
\end{equation*}

\vspace*{-2mm}
\begin{equation*}
\mathcal{B}_i\!=\!\sum\nolimits_{j=1}^{N}p_{ij}\hat{\nu}_{j}\!\left(\!B^{*}\! X_{\infty,j}^{c} B \!+\!R\right).
\end{equation*}
\end{small}
\vspace*{-4mm}

The optimal mode-dependent state feedback controller is given by 
{\small $u_k\!=\!K^{c}_{\theta_{k-1}}x_k$}, where 
{\small $K^{c}_{i} = \mathcal{B}_i^{-1} \mathcal{C}_i^{*}$}.
The performance index optimized by this controller is 
{\small $J^c\!=\!\limsup_{t\to\infty}\!\frac{1}{t}\mathbb{E}
(\sum_{k=0}^{t}(x_{k}^{*} Q x_{k}\!+\!u_{k}^{*}Ru_{k}) \mid \mathcal{G}_k)$}, which, 
for the optimal control law, equals to {\small $J^c_{\star}\!=\!\sum_{i=1}^{N}\mathbf{p}_i\mathrm{trace}(X_{\infty,i}^{c}\Sigma_w )$}.
%%%%%%%%%%%%%%%%%%%%%%%%%%%%%%%%%%%%%%%%%%%%%%%%%%%%%%%%%%%%%%%%%%%%%%%%%%%%%%%%
Clearly, the necessary condition for the existence of the stabilizing solution 
{\small $X_{\infty,i}^c$}, {\small $\forall i\!\leq\!N$}, to the CAREs \eqref{eq:care} 
is the (mean square) stabilizability of the system \eqref{eq:mjls}. However, 
for the best of our knowledge, such conditions have not been derived yet, since
e.g.~the well-known stabilizability conditions of the MJLSs with the 
instantaneous perfect observation of the operational mode 
\cite[pp. 57\,--\,58]{costa2006discrete} clearly do not account for the 
one-step delayed observation of the channel's state, 
as will be illustrated on a numerical example
in Section \ref{sec:ex}.
In the next subsection we derive the necessary and sufficient
conditions for the mean square stabilizability of the system \eqref{eq:mjls}.

%%%%%%%%%%%%%%%%%%%%%%%%%%%%%%%%%%%%%%%%%%%%%%%%%%%%%%%%%%%%%%%%%%%%%%%%%%%%%%%%
\section{Stabilizability analysis}\label{sec:stabilizability}
%%%%%%%%%%%%%%%%%%%%%%%%%%%%%%%%%%%%%%%%%%%%%%%%%%%%%%%%%%%%%%%%%%%%%%%%%%%%%%%%
Consider first the system \eqref{eq:mjls} without process noise, i.e.,

\vspace*{-2mm}
\begin{small}
\begin{equation}\label{eq:mjls_noiseless}
x_{k+1} \!=\! \left(A \!+\! \nu_{\theta_k} B K_{\theta_{k-1}}\right) x_k,
\end{equation}
\end{small}

\vspace*{-5.5mm}
%%%%%%%%%%%%%%%%%%%%%%%%%%%%%%%%%%%%%%%%%%%%%%%%%%%%%%%%%%%%%%%%%%%%%%%%%%%%%%%%
\noindent where {\small $K_{\theta_{k-1}}$} is a mode-dependent state-feedback 
controller with one time-step delayed operational mode observation, according to
the informational set {\small $\mathcal{G}_k$}. 
To account for the dependence on {\small $K_{\theta_{k-1}}$} in the behavior 
of the continuous state {\small $x_k$}, we augment the system's state with the 
memory of the previous state of the communication channel, i.e., the new aggregated 
state is {\small $\left(x_k, \nu_{\theta_k}, \theta_{k-1} \right)$}, or, equivalently, 
{\small $\left(x_k, \nu_{k}, \left(\theta_k, \theta_{k-1}\right) \right)$}, where the
last term underlines the similarity with the MCs with memory of order 
{\small $2$}, see \cite{wu2017markov} for additional details on MCs with memory.
The introduced memory is, however, fictitious, since the aggregated MC 
obeys to the Markov property of the memoryless chain {\small $\Theta$}:
%%%%%%%%%%%%%%%%%%%%%%%%%%%%%%%%%%%%%%%%%%%%%%%%%%%%%%%%%%%%%%%%%%%%%%%%%%%%%%%%

\vspace*{-4.5mm}
\begin{small}
\begin{align}
\mathrm{Pr}&\!\left(\theta_{k+1}\!=\!s_j, \theta_{k}\!=\!s_i \!\mid\! 
	\theta_{k}\!\neq\!s_i,\theta_{k-1}\!=\!s_{\ell}\right) = 0, \quad \,
	\forall s_{\ell} \!\in\!\mathbb{S}\label{eq:p_lij_0},\\
\mathrm{Pr}&\left(\theta_{k+1}\!=\!s_j, \theta_{k}\!=\!s_i \!\mid\! 
	\theta_{k}\!=\!s_i,\theta_{k-1}\!=\!s_{\ell}\right) = \nonumber \\
& \mathrm{Pr}\!\left(\theta_{k+1}\!=\!s_j \!\mid\! 
	\theta_{k}\!=\!s_i\right) = p_{ij},\quad \forall s_{\ell} \!\in\!\mathbb{S},\label{eq:p_lij_1}
\end{align}
\end{small}
%%%%%%%%%%%%%%%%%%%%%%%%%%%%%%%%%%%%%%%%%%%%%%%%%%%%%%%%%%%%%%%%%%%%%%%%%%%%%%%%

\vspace*{-5.5mm}
\noindent where the first equality in \eqref{eq:p_lij_1} is due to the fact that 
the intersection of a set with itself is the set itself, so
the joint probability
{\small $\mathrm{Pr}\!\left(\theta_{k}\!=\!s_i,\theta_{k}\!=\!s_i,
	\theta_{k-1}\!=\!s_{\ell}\right) \!=\! \mathrm{Pr}\!\left(\theta_{k}\!=\!s_i, 
	\theta_{k-1}\!=\!s_{\ell}\right)$}, which implies that 
{\small $\mathrm{Pr}\!\left(\theta_{k}\!=\!s_i \!\mid\! \theta_{k}\!=\!s_i,
	\theta_{k-1}\!=\!s_{\ell}\right) \!=\! 1$}, by
the definition of the conditional probability. %%%(as an axiom of probability), 
%%%%%%%%%%%%%%%%%%%%%%%%%%%%%%%%%%%%%%%%%%%%%%%%%%%%%%%%%%%%%%%%%%%%%%%%%%%%%%%%
Taken together with %%%general product rule, also known as 
the chain rule of 
probability, the last equality implies the first equality in \eqref{eq:p_lij_1}.
%%%%%%%%%%%%%%%%%%%%%%%%%%%%%%%%%%%%%%%%%%%%%%%%%%%%%%%%%%%%%%%%%%%%%%%%%%%%%%%% 
The second equality in \eqref{eq:p_lij_1} is clearly obtained from the direct 
application the Markov property.

%%%%%%%%%%%%%%%%%%%%%%%%%%%%%%%%%%%%%%%%%%%%%%%%%%%%%%%%%%%%%%%%%%%%%%%%%%%%%%%%
The joint probability of being in an augmented Markov state
{\small $\left(\theta_k, \theta_{k-1}\right)$} evolves according to 
\eqref{eq:p_lij_0} and \eqref{eq:p_lij_1}, so, after denoting 
{\small $\mathrm{Pr}\!\left(\theta_{k+1}\!=\!j, \theta_{k}\!=\!i\right)$} by 
{\small $\pi_{ij}^{(k+1)}$}, one has that

\vspace*{-1mm}
\begin{small}
\begin{equation}\label{eq:joint_probability}
\pi_{ij}^{(k+1)} \!=\! \sum\nolimits_{\ell=1}^{N}\pi_{\ell i}^{(k)}p_{ij}.
\end{equation}
\end{small}

\vspace*{-4mm}
%%%%%%%%%%%%%%%%%%%%%%%%%%%%%%%%%%%%%%%%%%%%%%%%%%%%%%%%%%%%%%%%%%%%%%%%%%%%%%%%
The joint probability {\small $\pi_{\ell i}^{(k)}$} may be defined through the
indicator function {\small $\mathbf{1}_{\left\{\theta_k=i,\theta_{k-1}=\ell \right\}}$}, 
that indicates the membership (or non-membership) of a given element in the set, 
as 

\vspace*{-2mm}
\begin{small}
\begin{equation}\label{eq:pi_il}
\mathbb{E}\!\left(\mathbf{1}_{\left\{\theta_k=i,\theta_{k-1}=\ell \right\}}\right)
\!=\!\pi_{\ell i}^{(k)}.
\end{equation}
\end{small}

\vspace*{-5.5mm}
%%%%%%%%%%%%%%%%%%%%%%%%%%%%%%%%%%%%%%%%%%%%%%%%%%%%%%%%%%%%%%%%%%%%%%%%%%%%%%%%
The indicator function {\small $\mathbf{1}_{\left\{\theta_k=i,\theta_{k-1}=\ell\right\}}$} 
allows us obtaining recursive difference equations for the first and second 
moments of the system's state, which are fundamental in deriving our result on 
stabilizability. Specifically, we define

\vspace*{-4mm}
%%%%%%%%%%%%%%%%%%%%%%%%%%%%%%%%%%%%%%%%%%%%%%%%%%%%%%%%%%%%%%%%%%%%%%%%%%%%%%%%
\begin{small}
\begin{equation}\label{eq:q}
\mathrm{m}_{\ell i}^{(k)}\!\triangleq\!\mathbb{E}\!\left(x_k
	\mathbf{1}_{\left\{\theta_k=i,\theta_{k-1}=\ell\right\}}\right),\quad
\mathbf{m}^{(k)}\!\triangleq\!\left[\mathrm{m}_{\ell i}^{(k)}\right]_{\ell,i=1}^{N},
\end{equation}
\end{small}

\vspace*{-9mm}
\begin{small}
\begin{equation}\label{eq:Q}
\mathrm{M}_{\ell i}^{(k)}\!\triangleq\!\mathbb{E}\!\left(x_k x_k^{*}
	\mathbf{1}_{\left\{\theta_k=i,\theta_{k-1}=\ell\right\}}\right),\quad
\mathbf{M}^{(k)}\!\triangleq\!\left[\mathrm{M}_{\ell i}^{(k)}\right]_{\ell,i=1}^{N},
\end{equation}
\end{small}

\vspace*{-4mm}
%%%%%%%%%%%%%%%%%%%%%%%%%%%%%%%%%%%%%%%%%%%%%%%%%%%%%%%%%%%%%%%%%%%%%%%%%%%%%%%%
\noindent so that the first and second moments of {\small $x_k$} are
%%%%%%%%%%%%%%%%%%%%%%%%%%%%%%%%%%%%%%%%%%%%%%%%%%%%%%%%%%%%%%%%%%%%%%%%%%%%%%%%

\vspace*{-3mm}
\begin{small}
\begin{equation}\label{eq:moments}
\mathbb{E}\!\left(x_{k}\right)\!=\!\!
	\sum\nolimits_{\ell=1}^{N}\!\sum\nolimits_{i=1}^{N}\!\!\mathrm{m}_{\ell i}^{(k)}\!,
	~~~
\mathbb{E}\!\left(x_{k}x_{k}^{*}\right)\!=\!\!
	\sum\nolimits_{\ell=1}^{N}\!\sum\nolimits_{i=1}^{N} \!\!\mathrm{M}_{\ell i}^{(k)}\!.
\end{equation}
\end{small}

\vspace*{-3mm}
%%%%%%%%%%%%%%%%%%%%%%%%%%%%%%%%%%%%%%%%%%%%%%%%%%%%%%%%%%%%%%%%%%%%%%%%%%%%%%%%
\begin{proposition}\label{prop:1}
Consider the system \eqref{eq:mjls_noiseless}. For all {\small $k\!\in\!\mathbb{N}_0$},
{\small $1\!\leq\!\ell,i,j\!\leq\!N$}, one has that

\vspace*{-2mm}
%%%%%%%%%%%%%%%%%%%%%%%%%%%%%%%%%%%%%%%%%%%%%%%%%%%%%%%%%%%%%%%%%%%%%%%%%%%%%%%%
\begin{small}
\begin{equation}\label{eq:evolution_qij}
\mathrm{m}_{ij}^{(k+1)}\!=\!\left(\!A\!\sum\nolimits_{\ell=1}^{N}\!\!\mathrm{m}_{\ell i}^{(k)}\!+\!
	B\!\sum\nolimits_{\ell=1}^{N}\!\!K_{\ell}\mathrm{m}_{\ell i}^{(k)}\hat{\nu}_{i}\!\right)
	\!p_{ij},
\end{equation}

\vspace*{-5mm}
\begin{align}\label{eq:evolution_Qij}
\mathrm{M}_{ij}^{(k+1)}\!=\!\Bigg(\!\!A\!\sum\nolimits_{\ell=1}^{N}\!&\mathrm{M}_{\ell i}^{(k)}
	\!A^{*}\!\!+\!B\!\sum\nolimits_{\ell=1}^{N}\!\!K_{\ell}\mathrm{M}_{\ell i}^{(k)}\!
	K_{\ell}^{*}\!B^{*}\!\hat{\nu}_i + \nonumber \\[-2mm]
	&2\Re\Bigg(\!\!A\!\!\sum\nolimits_{\ell=1}^{N}\!\mathrm{M}_{\ell i}^{(k)}\!
	K_{\ell}^{*}\!B^{*}\!\hat{\nu}_i\!\!\Bigg)\!\!\Bigg) p_{ij}.
\end{align}
\end{small}
\end{proposition}
%%%%%%%%%%%%%%%%%%%%%%%%%%%%%%%%%%%%%%%%%%%%%%%%%%%%%%%%%%%%%%%%%%%%%%%%%%%%%%%%
\vspace*{1mm}
\begin{proof}
See Appendix.
\end{proof}

%%%%%%%%%%%%%%%%%%%%%%%%%%%%%%%%%%%%%%%%%%%%%%%%%%%%%%%%%%%%%%%%%%%%%%%%%%%%%%%%
In the spirit of \cite{costa2006discrete}, Proposition~\ref{prop:1} allows us 
to define the operators {\small $\mathcal{L}\!\left(\cdot\right)\!\triangleq\!
\left[\mathcal{L}_{ij}\!\left(\cdot\right)\right]_{i,j=1}^{N}$} and 
{\small $\mathcal{T}\!\left(\cdot\right)\!\triangleq\!
\left[\mathcal{T}_{\ell i}\!\left(\cdot\right)\right]_{\ell,i=1}^{N}$,} both in 
{\small $\mathbb{B}\!\left(\mathbb{F}^{Nn_x\times Nn_x}\!\right)$,} as follows. 
{\small $\forall\mathbf{S}\!=\!\left[\mathrm{S}_{ij}\right]_{i,j=1}^{N}$,} 
{\small $\mathbf{T}\!=\!\left[\mathrm{T}_{\!ij}\right]_{i,j=1}^{N}$,} both in
{\small $\!\in\! \mathbb{F}^{Nn_x\times Nn_x}$,} we specify the 
{\it inner product} as 

\vspace*{-2mm}
\begin{small}
\begin{equation}\label{eq:innerProduct}
\left<\mathbf{S} ; \mathbf{T} \right> \!\triangleq\!
\sum\nolimits_{i=1}^{N}\sum\nolimits_{j=1}^{N}\mathrm{trace}\left(\mathrm{S}_{ij}^{*}\mathrm{T}_{\!ij}\right)\!\!\;,
\end{equation}
\end{small}

\vspace*{-5mm}
\noindent while the components of operators {\small $\mathcal{L}\!\left(\cdot\right)$}, 
{\small $\mathcal{T}\!\left(\cdot\right)$}, are defined by

\vspace*{-5mm}
\begin{small}
\begin{align}\label{eq:operatorLDefinition}
\mathcal{L}_{ij}\!\!\!\;\left(\mathbf{S}\right) \!\triangleq\!
\Bigg(\!\!&A\!\sum\nolimits_{\ell=1}^{N} \mathrm{S}_{\ell i}
	A^{*}\!+\!B\!\sum\nolimits_{\ell=1}^{N}\!K_{\ell}\mathrm{S}_{\ell i}
	K_{\ell}^{*}\!B^{*}\!\hat{\nu}_i + \nonumber \\[-3mm]
  & A\!\sum\nolimits_{\ell=1}^{N}\mathrm{S}_{\ell i}
	K_{\ell}^{*}\!B^{*}\!\hat{\nu}_i\!+\!
	B\!\sum\nolimits_{\ell=1}^{N}\!\!\,K_{\ell}\mathrm{S}_{\ell i}
	A^{*}\!\hat{\nu}_i\!\!
	\Bigg) p_{ij},
\end{align}
\end{small}

\vspace*{-8mm}
\begin{small}	
\begin{align}\label{eq:operatorTDefinition}	
\mathcal{T}_{\ell i}\!\!\!\;\left(\mathbf{S}\right) \!\triangleq\!
  &~ A^{*}\!\sum\nolimits_{j=1}^{N} p_{ij} \mathrm{S}_{ij}A\!+\!
     K_{\ell}^{*}B^{*}\!\sum\nolimits_{j=1}^{N} p_{ij} \mathrm{S}_{ij} B K_{\ell} \hat{\nu}_i
     + \nonumber \\[-.5mm]
  &~ K_{\ell}^{*}B^{*}\!\sum\nolimits_{j=1}^{N} p_{ij} \mathrm{S}_{ij} A \hat{\nu}_i \!+\!
     A^{*}\!\sum\nolimits_{j=1}^{N} p_{ij} \mathrm{S}_{ij} K_{\ell} \hat{\nu}_i.
\end{align}
\end{small}
\vspace*{-5mm}

\begin{remark}\label{rem:operators}
Clearly, we have that {\small $\left(\mathcal{L}\!\left(\mathbf{S}\right)\right)^{*}\!=\!\mathcal{L}\!\left(\mathbf{S}^{*}\right)$,}
and it is immediate to verify (starting from \eqref{eq:innerProduct}, applying \eqref{eq:operatorLDefinition}, \eqref{eq:operatorTDefinition}, 
linearity of the trace operator and its invariance under the cyclic permutations) that {\small $\mathcal{T}\!\left(\cdot\right)$} is the {\it adjoint} operator of
{\small $\mathcal{L}\!\left(\cdot\right)$}, i.e., {\small $\mathcal{L}^{*}\!=\!\mathcal{T}$.} 
This is a generalization of \cite[Prop.~3.2, p.~33]{costa2006discrete}. 
Furthermore, it is evident from their definitions \eqref{eq:operatorLDefinition}, \eqref{eq:operatorTDefinition} that 
{\small $\mathcal{L}\!\left(\cdot\right)$} and {\small $\mathcal{T}\!\left(\cdot\right)$} are Hermitian and positive operators.
%i.e., they map Hermitian matrices into Hermitian matrices and PSD matrices into PSD ones.
%%%(i.e., they map $\succeq$
\end{remark}
%%%\begin{proof}
%%%See Appendix.
%%%\end{proof}
%%%%%%%%%%%%%%%%%%%%%%%%%%%%%%%%%%%%%%%%%%%%%%%%%%%%%%%%%%%%%%%%%%%%%%%%%%%%%%%%
\begin{small}
\begin{equation*}
\text{Define}\quad\Delta\mathrm{P}_{1}\!\triangleq\!\left(\bigconcatenate\limits_{j=1}^{N}\!
	\left(\bigoplus\limits_{i=1}^{N} p_{ij}\!\! \right)\!\!\right)'\!,\quad
\Delta\mathrm{P}_{\hat{\nu}}\!\triangleq\!\!
	\left(\bigconcatenate\limits_{j=1}^{N}\!
	\left(\bigoplus\limits_{i=1}^{N}\hat{\nu}_i p_{ij}\!\!\right)
	\!\!\right)'\!.
\end{equation*}
\end{small}

\vspace*{-3mm}
%%%%%%%%%%%%%%%%%%%%%%%%%%%%%%%%%%%%%%%%%%%%%%%%%%%%%%%%%%%%%%%%%%%%%%%%%%%%%%%%
\indent Then, the matrix forms of \eqref{eq:evolution_qij} and of 
\eqref{eq:evolution_Qij} can be written respectively as

\vspace*{-2mm}
%%%%%%%%%%%%%%%%%%%%%%%%%%%%%%%%%%%%%%%%%%%%%%%%%%%%%%%%%%%%%%%%%%%%%%%%%%%%%%%%
\begin{small}
\begin{equation}\label{eq:vec2q}
\mathrm{vec}^{2}\!\left(\mathbf{m}^{(k+1)}\right)\!=\!
	\bm{\Psi} \mathrm{vec}^{2}\!\left(\mathbf{m}^{(k)}\right)\!,
\end{equation}

\vspace*{-3mm}
\begin{equation*}
\bm{\Psi}\!=\!
	\Delta\mathrm{P}_{\hat{\nu}}\otimes\!
	\left(\bigconcatenate\limits_{j=1}^{N}\!\left(BK_{j}\right)\!\right)\!+
	\Delta\mathrm{P}_{1}\otimes\!\left(\bigconcatenate\limits_{j=1}^{N}A\!\right)\!,
\end{equation*}

\vspace*{-1mm}
\begin{equation}\label{eq:vec2Q}
\mathrm{vec}^{2}\!\left(\mathbf{M}^{(k+1)}\right)\!=\!
	\bm{\Lambda} \mathrm{vec}^{2}\!\left(\mathbf{M}^{(k)}\right),
\end{equation}

\vspace*{-4mm}
\begin{align}\label{eq:Lambda}
\bm{\Lambda}\!=&
	\Delta\mathrm{P}_{\hat{\nu}}\otimes\!
	\left(\bigconcatenate\limits_{j=1}^{N}\!\bigg(\!\!\!\Big(\!\!\!\left(
	\bar{B}\bar{K}_{j}\right)\!\otimes\!\left(BK_{j}\right)\!\!\Big)\!\!+\!2
	\Re\Big(\!\!\!\left(\bar{B}\bar{K}_{j}\right)\!\otimes\!A\Big)\!\!\bigg)
	\!\!\!\right)\!\!+\!\nonumber\\[-3mm]
	&\qquad\qquad\quad\Delta\mathrm{P}_{1}\otimes\!\left(\bigconcatenate\limits_{j=1}^{N}\!
	\left(\bar{A}\!\otimes\!A\right)\!\!\right)\!.
\end{align}
\end{small}
%%%%%%%%%%%%%%%%%%%%%%%%%%%%%%%%%%%%%%%%%%%%%%%%%%%%%%%%%%%%%%%%%%%%%%%%%%%%%%%%

\vspace*{-2mm}
\begin{proposition}\label{prop:rho_Lambda_Psi}
If {\small $\rho\!\left({\bm \Lambda}\right)\!<\!1$} then 
{\small $\rho\!\left({\bm \Psi}\right)\!<\!1$}.
\end{proposition}
\begin{proof}
See Appendix.
\end{proof}

From \eqref{eq:evolution_Qij} and \eqref{eq:operatorLDefinition}, it is immediate to verify that
{\small $\mathbf{M}^{(k+1)}\!=\!\mathcal{L}(\mathbf{M}^{(k)})$},
and {\small $\forall \mathbf{S}\!=\!\left[\mathrm{S}_{ij}\right]_{i,j=1}^{N}\!\in\!
\mathbb{F}^{Nn_x\times Nn_x}$}, {\small $\mathrm{S}_{ij}\!\succeq\!0$}, by 
\eqref{eq:operatorLDefinition}, and
\eqref{eq:operatorTDefinition} together with Remark~\ref{rem:operators}, we have that 
{\small $\mathrm{vec}^{2}(\mathcal{L}(\mathbf{S})) \!=\!
{\bm \Lambda}\mathrm{vec}^{2}(\mathbf{S})$,
$\mathrm{vec}^{2}(\mathcal{T}(\mathbf{S})) \!=\! {\bm \Lambda}^{\!\!*}\mathrm{vec}^{2}(\mathbf{S})$}.
Thus, we have also that {\small $\rho\!\left(\mathcal{T}\right)\!=\!\rho\!\left(\mathcal{L}\right)\!=\!\rho\!\left({\bm \Lambda}\right)$.}

\begin{definition}\label{def:stabilizability}
A system \eqref{eq:mjls} is \textit{mean square stabilizable} if for any initial 
condition {\small $\left(\hat{x}_0,\theta_0\right)$} there exist a mode-dependent 
state-feedback controller {\small $\mathbf{K}\!\triangleq\!\left(K_i\right)_{i=1}^{N}$}
with one time-step delayed operational mode observation, such that the system 
\eqref{eq:mjls} is mean square stable.
\end{definition}

In order to apply the usual definition of the mean square stability
\cite[pp. 36--37]{costa2006discrete} to the system \eqref{eq:mjls}, we denote the 
operational modes of the system \eqref{eq:mjls_noiseless} by 
{\small $\varphi_k\!\triangleq\!\left(\nu_{\theta_k},\theta_{k-1}\right)$}, which is a
{\small $2N^2$}-ary random quantity. Then, \eqref{eq:mjls} becomes

\vspace*{-2mm}
%%%%%%%%%%%%%%%%%%%%%%%%%%%%%%%%%%%%%%%%%%%%%%%%%%%%%%%%%%%%%%%%%%%%%%%%%%%%%%%%
\begin{small}
\begin{equation}\label{eq:mjls_alternative}
x_{k+1} \!=\! \mathfrak{A}_{\varphi_k} x_k \!+\!w_k,
\end{equation}
\end{small}
%%%%%%%%%%%%%%%%%%%%%%%%%%%%%%%%%%%%%%%%%%%%%%%%%%%%%%%%%%%%%%%%%%%%%%%%%%%%%%%%

\vspace*{-5mm}
\noindent where 
{\small $\mathfrak{A}_{\varphi_k}\!\triangleq\!A\!+\!\nu_{\theta_k}BK_{\theta_{k-1}}$}, 
so considering each possible value of {\small $\varphi_k$}, we obtain also
{\small $\bm{\mathfrak{A}}\triangleq\!\left(\mathfrak{A}_i\right)_{i=1}^{2N^2}$}. Then, we
can recall the usual definition of mean square stability of MJLSs:
\begin{definition}\label{def:mss}
An MJLS \eqref{eq:mjls_alternative} is \textit{mean square stable} if for any 
initial condition {\small $\left(\hat{x}_0,\hat{\varphi}_0\right)$} there exist equilibrium points
{\small ${x}_e$} and {\small ${M}_e$} (independent from initial conditions 
{\small $\hat{x}_0$} and {\small $\hat{\varphi}_0$}), such that
%%%%%%%%%%%%%%%%%%%%%%%%%%%%%%%%%%%%%%%%%%%%%%%%%%%%%%%%%%%%%%%%%%%%%%%%%%%%%%%%

\vspace*{-5mm}
\begin{small}
\begin{equation}\label{eq:mss_general}
\lim_{k \to \infty}\left\|\mathbb{E}\!\left(
	x_k \right)\!-\!{x}_e\right\|\!=\!0, \quad
\lim_{k \to \infty}\left\|\mathbb{E}\!\left(
	x_k x_k^* \right)\!-\!{M}_{e} \right\| \!=\! 0.
\end{equation}
\end{small}
\end{definition}

\vspace*{1mm}
%%%%%%%%%%%%%%%%%%%%%%%%%%%%%%%%%%%%%%%%%%%%%%%%%%%%%%%%%%%%%%%%%%%%%%%%%%%%%%%%
It is worth mentioning \cite[p. 37, Remark 3.10]{costa2006discrete} that in 
noiseless case, i.e., when {\small $w_k\!=\!0$} in \eqref{eq:mjls_alternative}, the 
conditions \eqref{eq:mss_general} defining mean square stability become

\vspace*{-2mm}
\begin{small}
\begin{equation} \label{eq:mss_noiseless}
   \lim_{k \to \infty} \mathbb{E}\!\left( x_k \right) = 0, \qquad
   \lim_{k \to \infty} \mathbb{E}\!\left( x_k x_k^* \right) = 0    
\end{equation}
\end{small}

\vspace*{-5mm}
%%%%%%%%%%%%%%%%%%%%%%%%%%%%%%%%%%%%%%%%%%%%%%%%%%%%%%%%%%%%%%%%%%%%%%%%%%%%%%%%
\begin{proposition}\label{prop:delayedNoiselessMeanSquareStability}
The system \eqref{eq:mjls_noiseless}, i.e., \eqref{eq:mjls_alternative} 
with {\small $w_k\!=\!0$} for all {\small $k\!\in\!\mathbb{N}_0$}, is 
mean square stable if and only if {\small $\rho\!\left({\bm \Lambda}\right)\!<\!1$}.
\end{proposition}
\begin{proof}
See Appendix.
\end{proof}

%%%%%%%%%%%%%%%%%%%%%%%%%%%%%%%%%%%%%%%%%%%%%%%%%%%%%%%%%%%%%%%%%%%%%%%%%%%%%%%%
\begin{proposition}\label{prop:delayedNoiselessLyapunov}
Consider the system \eqref{eq:mjls_noiseless}. 
Then {\small $\rho\!\left({\bm \Lambda}\right)\!<\!1$} if and only if for any
{\small $\mathbf{Z}\!=\!\!\left[\mathrm{Z}_{ij}\right]_{i,j=1}^{N}\!\in\!
\mathbb{F}^{Nn_x\times Nn_x}$}, {\small $\mathrm{Z}_{ij}\!\succ\!0$}, there exists a unique
{\small $\mathbf{Y}\!=\!\!\left[\mathrm{Y}_{ij}\right]_{i,j=1}^{N}\!\in\!
\mathbb{F}^{Nn_x\times Nn_x}$}, {\small $\mathrm{Y}_{ij}\!\succ\!0$}, such that

\vspace*{-2mm}
\begin{small}
\begin{equation}\label{eq:Lyapunov}
\mathbf{Y} \!-\! \mathcal{L}\!\left(\mathbf{Y}\right) \!=\! \mathbf{Z}.
\end{equation}
\end{small}
\end{proposition}
\begin{proof}
See Appendix.
\end{proof}

%%%%%%%%%%%%%%%%%%%%%%%%%%%%%%%%%%%%%%%%%%%%%%%%%%%%%%%%%%%%%%%%%%%%%%%%%%%%%%%%
\begin{proposition}\label{prop:noisyQ}
Consider the system \eqref{eq:mjls_alternative}, where {\small $w_k$} is a Gaussian 
white process noise with zero mean and covariance matrix {\small $\Sigma_w$}, 
assumed to be independent from the initial state {\small $x_0$} and the stochastic 
variables {\small $\nu_{k}$}, {\small $\theta_{k}$} and {\small $\theta_{k-1}$}, 
{\small $\forall k\!\in\!\mathbb{N}_{0}$}. Then, for all 
{\small $1\!\leq\!\ell,i,j\!\leq\!N$}, one has that \eqref{eq:evolution_qij}
still holds, and

\vspace*{-5mm}
%%%%%%%%%%%%%%%%%%%%%%%%%%%%%%%%%%%%%%%%%%%%%%%%%%%%%%%%%%%%%%%%%%%%%%%%%%%%%%%%
\begin{small}
\begin{align}\label{eq:evolution_noisy_Qij}
&\mathrm{M}_{ij}^{(k+1)}\!=\!\Bigg(\!\!A\!\!\sum\nolimits_{\ell=1}^{N}\mathrm{M}_{\ell i}^{(k)}
	\!A^{*}\!\!+\!B\!\sum\nolimits_{\ell=1}^{N}K_{\ell}\mathrm{M}_{\ell i}^{(k)}\!
	K_{\ell}^{*}\!B^{*}\!\hat{\nu}_i + \nonumber \\[-1.5mm]
	&\quad~~2\Re\Bigg(\!\!A\!\!\sum\nolimits_{\ell=1}^{N}\mathrm{M}_{\ell i}^{(k)}\!
	K_{\ell}^{*}\!B^{*}\!\hat{\nu}_i\!\!\Bigg)\!\!\Bigg) p_{ij} + 
	\Sigma_w\!\!\sum\nolimits_{\ell=1}^{N}\pi_{\ell i}^{(k)}p_{ij}.
\end{align}
\end{small}
\end{proposition}
\vspace*{1mm}
\begin{proof}
See Appendix.
\end{proof}

%%%%%%%%%%%%%%%%%%%%%%%%%%%%%%%%%%%%%%%%%%%%%%%%%%%%%%%%%%%%%%%%%%%%%%%%%%%%%%%%
After noting that the last addend in \eqref{eq:evolution_noisy_Qij} 
can be written as {\small $\Sigma_w\!\!\sum\limits_{\ell=1}^{N}\!\left(\!
\pi_{\ell i}^{(k)}\!\!\otimes\!\mathbb{I}_{n_x}\!\right)\!\mathbb{I}_{n_x}p_{ij}$},
we define {\small ${\bm \Pi}^{(k)}\!\triangleq\!\Big[\pi_{\ell i}^{(k)}\!\!\otimes\!
\mathbb{I}_{n_x}\Big]_{\ell, i=1}^{N}$}, so that the matrix form of 
\eqref{eq:evolution_noisy_Qij} is given by

\vspace*{-1mm}
%%%%%%%%%%%%%%%%%%%%%%%%%%%%%%%%%%%%%%%%%%%%%%%%%%%%%%%%%%%%%%%%%%%%%%%%%%%%%%%%
\begin{small}
\begin{equation}\label{eq:vec2S_noisy}
\mathrm{vec}^{2}\!\!\left(\!\mathbf{M}^{(k+1)}\!\right)\!=\!
	\bm{\Lambda}\mathrm{vec}^{2}\!\!\left(\!\mathbf{M}^{(k)}\!\right) \!+\!
	{\bm \Upsilon}
	\mathrm{vec}^{2}\!\left(\!{\bm \Pi}^{(k)}\!\right),
\end{equation}

\vspace*{-3mm}
\begin{equation}\label{eq:Upsilon}
{\bm \Upsilon} \!=\!
\Delta\mathrm{P}_{1}\otimes\!\left(\bigconcatenate\limits_{j=1}^{N}
	\left(\mathbb{I}_{n_x}\!\!\otimes\!\Sigma_w\!\right)\!\right)\!.
\end{equation}
\end{small}

\vspace*{-3mm}
%%%%%%%%%%%%%%%%%%%%%%%%%%%%%%%%%%%%%%%%%%%%%%%%%%%%%%%%%%%%%%%%%%%%%%%%%%%%%%%%
\begin{theorem}\label{theorem:delayedMeanSquareStability}
Consider the system \eqref{eq:mjls_alternative}, where {\small $w_k$} is Gaussian 
white process noise with zero mean and covariance matrix {\small $\Sigma_w$}, 
assumed to be independent from the initial state $x_0$ and the stochastic variables 
{\small $\nu_{k}$}, {\small $\theta_{k}$} and {\small $\theta_{k-1}$}, 
{\small $\forall k\!\in\!\mathbb{N}_{0}$}. Also assume that MC 
{\small $\Theta$} representing the evolution of the Markov 
channel is ergodic. Then the system is mean square stable if and only 
if {\small $\rho\!\left({\bm \Lambda}\right)\!<\!1$}.
\begin{proof}
See Appendix.
\end{proof}
\end{theorem}

%%%%%%%%%%%%%%%%%%%%%%%%%%%%%%%%%%%%%%%%%%%%%%%%%%%%%%%%%%%%%%%%%%%%%%%%%%%%%%%%
\begin{theorem}\label{theorem:delayedMeanSquareStabilizability}
A system \eqref{eq:mjls} is mean square stabilizable if and only if 
there are
{\small ${\mathbf{V}_{_{\!1}}}\!=\!\big[V_{{_{\!1}}{\ell i}}\big]_{\ell,i=1}^{N}$},
{\small ${\mathbf{V}_{_{\!2}}}\!=\!\big[V_{{_{\!2}}{\ell i}}\big]_{\ell,i=1}^{N}$},
and {\small ${\mathbf{V}_{_{\!3}}}\!=\!\big[V_{{_{\!3}}{\ell i}}\big]_{\ell,i=1}^{N}$},
{\small $\mathbf{L}\!=\!\big[L_{i}\big]_{i=1}^{N}$},
where {\small $V_{{_{\!1}}{\ell i}}\!\in\!\mathbb{F}^{n_{x}\times n_{x}}$},
{\small $V_{{_{\!1}}{\ell i}}\!\succ\!0$},
{\small $V_{{_{\!2}}{\ell i}}\!\in\!\mathbb{F}^{n_{x}\times n_{u}}$}, 
{\small $V_{{_{\!3}}{\ell i}}\!\in\!\mathbb{F}^{n_{u}\times n_{u}}$}, 
{\small $V_{{_{\!3}}{\ell i}}\!\succeq\!0$}, and
{\small $L_{i}\!\in\!\mathbb{F}^{n_{u}\times n_{x}}$},
such that {\small $\forall i,j$}, one has that

\vspace*{-5mm}
%%%%%%%%%%%%%%%%%%%%%%%%%%%%%%%%%%%%%%%%%%%%%%%%%%%%%%%%%%%%%%%%%%%%%%%%%%%%%%%%
\begin{small}
\begin{subequations}\label{eq:delayedStabilizabilityLMI}
\begin{equation}
\sum\limits_{\ell=1}^{N}p_{ij}\!\left(AV_{{_{\!1}}{\!\ell i}}A^{*}\!\!+\!
AV_{{_{\!2}}{\ell i}}B^{*}\!\hat{\nu}_{i}\!+\!BV^{*}_{{_{\!2}}{\ell i}}A^{*}\!\hat{\nu}_{i}\!+\!
BV_{{_{\!3}}{\ell i}}B^{*}\!\hat{\nu}_{i}\right)-V_{{_{\!1}}{i j}}\!\prec\!0, \nonumber
\end{equation}

%%%%%%%%%%%%%%%%%%%%%%%%%%%%%%%%%%%%%%%%%%%%%%%%%%%%%%%%%%%%%%%%%%%%%%%%%%%%%%%%
\vspace*{-5mm}
\begin{equation}\label{eq:delayedStabilizabilityLMI_b}
\begin{bmatrix}
V_{{_{\!1}}{i j}} & V_{{_{\!2}}{i j}} \\
V^{*}_{{_{\!2}}{i j}} & L_{i}V_{{_{\!2}}{i j}}
\end{bmatrix}\!\succeq\!0,
\end{equation}
\begin{equation}\label{eq:delayedStabilizabilityLMI_c}
V_{{_{\!3}}{i j}} \!\succeq\! L_{i}V_{{_{\!2}}{i j}}
\end{equation}
%\begin{tabular}{ll}
%\begin{minipage}[c]{.38\linewidth}
%\begin{equation}\label{eq:delayedStabilizabilityLMI_b}
%\!\!\!\!\!\!\!\!\!\begin{bmatrix}
%V_{{_{\!1}}{i j}} & V_{{_{\!2}}{i j}} \\
%V^{*}_{{_{\!2}}{i j}} & L_{i}V_{{_{\!2}}{i j}}
%\end{bmatrix}\!\succeq\!0,\!\!\!\!\!\!\!\!\!\!\!\!\!
%\end{equation}
%\end{minipage}
%& 
%\begin{minipage}[c]{.46\linewidth}
%\begin{equation}\label{eq:delayedStabilizabilityLMI_c}
%~~~~~~~~~V_{{_{\!3}}{i j}} \!\succeq\! L_{i}V_{{_{\!2}}{i j}}
%\end{equation}
%\end{minipage}
%\end{tabular}
\end{subequations}
\end{small}
\end{theorem}
\begin{proof}
See Appendix.
\end{proof}

\textbf{Mode-independent control.}
%%%%%%%%%%%%%%%%%%%%%%%%%%%%%%%%%%%%%%%%%%%%%%%%%%%%%%%%%%%%%%%%%%%%%%%%%%%%%%%%
If {\small $X_{\infty,i}^{c}\!=\!\hat{X}_{\infty}^{c}$, $\forall i\!\leq\!N$}, 
the solution to the CAREs \eqref{eq:care} is mode-independent. Although 
mode-independent control is obviously more conservative than mode-dependent, it 
is very appealing in several scenarios, since it avoids the necessity to implement 
the online measurements of the Markov channel state.

\begin{theorem}\label{theorem:solModifiedARE}
Let $\hat{\nu}\!=\!\sum_{i=1}^{N}\mathbf{p}_i\hat{\nu}_i$. Then, the solution to the MARE 
provides the mode-independent solution to the CAREs.
\begin{proof}
See Appendix.
\end{proof}
\end{theorem}
So, the performance index for mode-independent solution to CAREs \eqref{eq:care}
is given exactly by {\small $J^b_{\star}$}. We stress that the mode-independent
solution to CARE implicitly requires the system \eqref{eq:mjls} to be \textit{strongly mean square 
stabilizable}, i.e., to satisfy requirement \eqref{eq:delayedStabilizabilityLMI}. 
%%%%%%%%%%%%%%%%%%%%%%%%%%%%%%%%%%%%%%%%%%%%%%%%%%%%%%%%%%%%%%%%%%%%%%%%%%%%%%%%
In the next section we show on a numerical example that when a networked control 
system is stabilizable, the controllers based on the mode-dependent and mode-independent 
solutions to CAREs are stabilizing, and the difference between their performance 
indices is small, while when the system is not stabilizable, the controllers obviously
cannot stabilize the system, even when {\small $\hat{\nu}\!>\nu_c$}.

%%%%%%%%%%%%%%%%%%%%%%%%%%%%%%%%%%%%%%%%%%%%%%%%%%%%%%%%%%%%%%%%%%%%%%%%%%%%%%%%
\section{NUMERICAL EXAMPLE}\label{sec:ex}
%%%%%%%%%%%%%%%%%%%%%%%%%%%%%%%%%%%%%%%%%%%%%%%%%%%%%%%%%%%%%%%%%%%%%%%%%%%%%%%%
%2345678901234567890123456789012345678901234567890123456789012345678901234567890
%        1         2         3         4         5         6         7         8
%%%%%%%%%%%%%%%%%%%%%%%%%%%%%%%%%%%%%%%%%%%%%%%%%%%%%%%%%%%%%%%%%%%%%%%%%%%%%%%%
Consider the inverted pendulum on a cart as in \cite{franklin1994feedback}.
The cart's mass is $0.5$kg, while the pendulum has mass of $0.2$kg, and inertia 
about its mass center of $0.006$kg$\cdot$m$^2$; the distance from the pivot to 
the pendulum's mass center is $0.3$m, the coefficient of friction for cart is $0.1$.
The state variables are the cart's position coordinate {\small $\mathrm{x}$} 
and pendulum's angle from vertical {\small $\phi$}, together with respective first 
derivatives. We aim to design a controller that stabilizes the 
pendulum in up-right position, corresponding to unstable equilibrium point 
{\small $\mathrm{x}^{\star}\!=\!0\,$}m, {\small $\phi^{\star}\!=\!0\,$}rad, so the system state is
defined by {\small $x\!=\!\begin{bmatrix}
\delta\mathrm{x}, \delta\dot{\mathrm{x}}, \delta\phi, \delta\dot{\phi}
\end{bmatrix}'$}, where 
{\small $\delta\mathrm{x}(t)\!=\!\mathrm{x}(t)\!-\!\mathrm{x}^{\star}$}, and
{\small $\delta\phi(t)\!=\!\phi(t)\!-\!\phi^{\star}$}. The initial state 
{\small $x_0\!=\!\begin{bmatrix} 0, 0, \frac{\pi}{10}, 0 \end{bmatrix}'\!$}.
The state space model of the system is linearized around the unstable 
equilibrium point and discretized with sampling period 
{\small $\mathrm{T}_{\mathrm{s}}\!=\!0.01\,$}s (see Section \ref{sec:whart}; 
the lagging effect associated with a zero-order hold discretisation is neglected).
%is
%
%\vspace*{-8mm}
%%%%%%%%%%%%%%%%%%%%%%%%%%%%%%%%%%%%%%%%%%%%%%%%%%%%%%%%%%%%%%%%%%%%%%%%%%%%%%%%%
%\begin{small}
%\begin{equation*}
%A\!=\!\begin{bmatrix}
%	1 &  0.099 & 0.014 & 0     \\
%	0 &  0.982 & 0.279 & 0.014 \\
%	0 & -0.002 & 1.160 & 0.105 \\
%	0 & -0.047 & 3.280 & 1.160
%\end{bmatrix}\!,\,
%B\!=\!\begin{bmatrix}
%	0.009 \\ 0.182 \\ 0.023 \\ 0.474
%\end{bmatrix}\!,
%\end{equation*}
%\end{small}
%%%%%%%%%%%%%%%%%%%%%%%%%%%%%%%%%%%%%%%%%%%%%%%%%%%%%%%%%%%%%%%%%%%%%%%%%%%%%%%%%
%
%\vspace*{-3mm}
%\noindent 
The weighting matrices are {\small $Q\!=\!\bigoplus (5000,0,100,0)$, $R\!=\!1$},
while the process noise is characterized by the covariance matrix
{\small $\Sigma_w\!=\!v v^{*}$},
{\small $v\!=\!\begin{bmatrix}0.030, 0.100, 0.010, 0.150\end{bmatrix}'\!$}.
The state matrix {\small $A$} is unstable, since it has an eigenvalue {\small $1.058$}, but
it is easy to verify that {\small $R\!\succ\!0$}, {\small $Q\!\succeq\!0$}, and
the pairs {\small $\left(A,B\right)$} and 
{\small $\left(A,Q\right)$} are controllable, so the closed-loop system 
is asymptotically stable, if {\small $\nu_k\!=\!1$} {\small $\forall k$}. 
Note that the critical probability {\small $\nu_c$} for the networked control over 
Bernoulli channel for this system is {\small $0.106$}.

%%%%%%%%%%%%%%%%%%%%%%%%%%%%%%%%%%%%%%%%%%%%%%%%%%%%%%%%%%%%%%%%%%%%%%%%%%%%%%%%
Consider the WirelessHART channel with two users characterized by the same 
channel hopping sequence (which is the worst possible scenario that also 
accounts for malicious behaviors such as deliberate jamming), typical value 
of transmitter and interferer power, $0\!$~dBm, the distance between 
the receiver and transmitter of interest of $10\,$m, and the distance between the 
receiver and interferer of {\small $d_{i}\,$}m. 

To account for strongly stabilizable scenario, let the distance to interferer be  
{\small $\hat{d}_{i}\!=\!14\,$}m. This produces the models
described in Section~\ref{sec:whart},
{\small $\hat{\Gamma} \sim\mathcal{N}\!\left(\hat{\mu},\hat{\sigma}^2\right)$}, with
{\small $\hat{\mu}\!=\!10.15\!$}~dB, {\small $\hat{\sigma}\!=\!4.85\!$}~dB, for which
{\small $\mathbb{E}(\mathrm{R}_{\mathrm{p}}(\hat{\Gamma}))\!=\!1\!-\!\hat{\nu}\!=0.008$},
so that for
{\small $\varepsilon^{\star}\!=\!3.17\cdot10^{-10}$} we get
{\small $\hat{\mathrm{L}}_{\mathrm{B}}\!\left(\varepsilon^{\star}\right)\!=\!11$}.
The same values of {\small $\mathrm{L}_{\mathrm{B}}$} and average PER are 
achieved by a Gilbert channel having {\small $\Gamma^{\star}$}
as the threshold that partitions the range of SNIR.
The related MJLS is strongly mean square stabilizable, 
since the optimal linear quadratic regulation with Bernoulli dropouts has 
{\small $\rho(\hat{\bm \Lambda}^{b})\!=\!0.909$} (and
{\small $\hat{J}_{\star}^{b}\!=\!316.663$}). The mode-dependent optimal Markovian controller 
with one time-step delayed mode observations has 
{\small $\rho(\hat{\bm {\Lambda}}^{c})\!=\!0.909$} and {\small $\hat{J}_{\star}^{c}\!=\!316.619$}.
So, the mode-independent solution is very appealing in this case.

If the interferer closes the distance to {\small $\check{d}_{i}\!=\!3.50\!$}~m, the analytic model 
becomes {\small $\check{\Gamma}$} having
{\small $\check{\mu}\!=\!-5.22\!$}~dB, 
{\small $\check{\sigma}\!=\!4.87\!$}~dB,
{\small $\mathbb{E}(\mathrm{R}_{\mathrm{p}}(\check{\Gamma}))\!=\!0.767$}, and
% and
{\small $\check{\mathrm{L}}_{\mathrm{B}}\!\left(\varepsilon^{\star}\right)\!=\!975$}.
The Gilbert channel is still able to track this behavior. 
In this case, the related MJLS is mean square stabilizable, but not strongly mean square stabilizable, 
since the optimal state feedback controller under TCP-like protocols has 
{\small $\rho(\check{\bm \Lambda}^{b})\!=\!1.001$}, and
{\small $\check{J}_{\star}^{b}\!=\!1186124.787$}, while the mode-dependent optimal Markovian controller 
with one time-step delayed mode observations still has 
{\small $\rho(\check{\bm {\Lambda}}^{c})\!=\!0.964$}, and {\small $\check{J}_{\star}^{c}\!=\!497.512$}.

If the interferer reaches the distance {\small $\tilde{d}_{i}\!=\!2.63\!$}~m, then the analytic model becomes
{\small $\tilde{\Gamma}$} having
{\small $\tilde{\mu}\!=\!-7.70\!$}~dB, 
{\small $\tilde{\sigma}\!=\!4.87\!$}~dB,
{\small $\mathbb{E}(\mathrm{R}_{\mathrm{p}}(\tilde{\Gamma}))\!=\!0.891$},
% and
{\small $\tilde{\mathrm{L}}_{\mathrm{B}}\!\left(\varepsilon^{\star}\right)\!=\!3719.$} 
According to the classical stabilizability conditions neglecting the one-time step mode observation delay \cite[pp.\,57\,--\,58]{costa2006discrete},
% % which do not account for one time-step delayed mode observations, 
the system is % mean square
stabilizable (with the associated spectral radius of {\small 0.999}), while in reality it is not:
% % , see Theorem~\ref{theorem:delayedMeanSquareStabilizability}.
%In fact, 
a tentative application of the optimal Markovian controller gives
{\small $\rho(\tilde{\bm {\Lambda}}^{c})\!=\!1.058$} and {\small $\tilde{J}_{\star}^{c}\!=\!502942.379$}.
Notably, {\small $1\!-\!\mathbb{E}(\mathrm{R}_{\mathrm{p}}(\tilde{\Gamma}))\!=\!0.109\!>0.106\!=\!\nu_c$}:
an abstraction of {\small $\tilde{\Gamma}$} with a Bernoulli channel would produce the misleading results.

\begin{figure}
	\centering
	\includegraphics[width=0.485\textwidth]{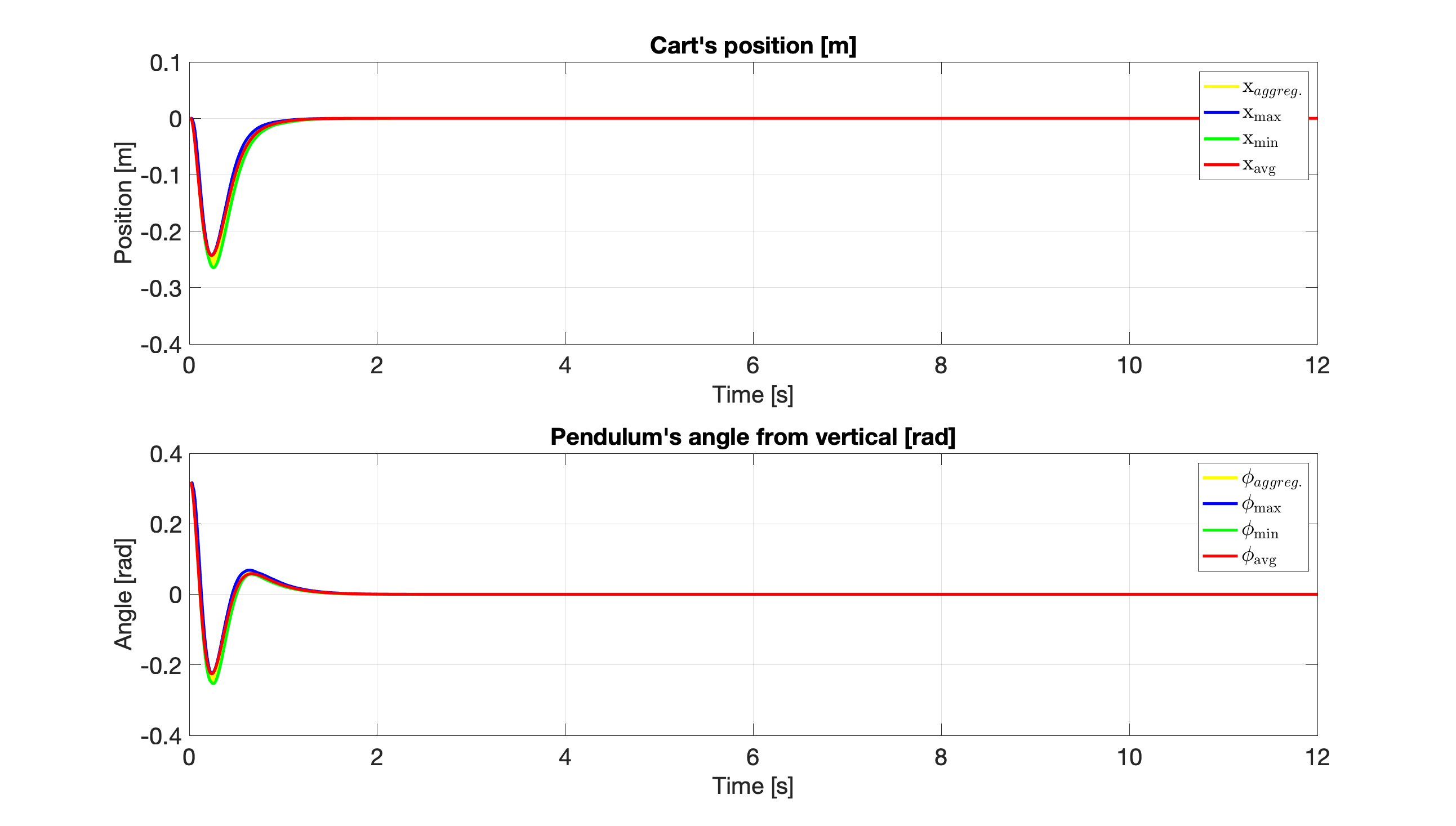}\\
	\vspace{-4mm}	
	\caption{Traces of the system's state that are generated under the Markovian control law over WirelessHART channel {\small $\hat{\Gamma}$} having {\small $\hat{d}_{i}\!=\!14\,$}m.}
	\label{fig:markov-14}
\end{figure}

\begin{figure}
	\centering
	\includegraphics[width=0.485\textwidth]{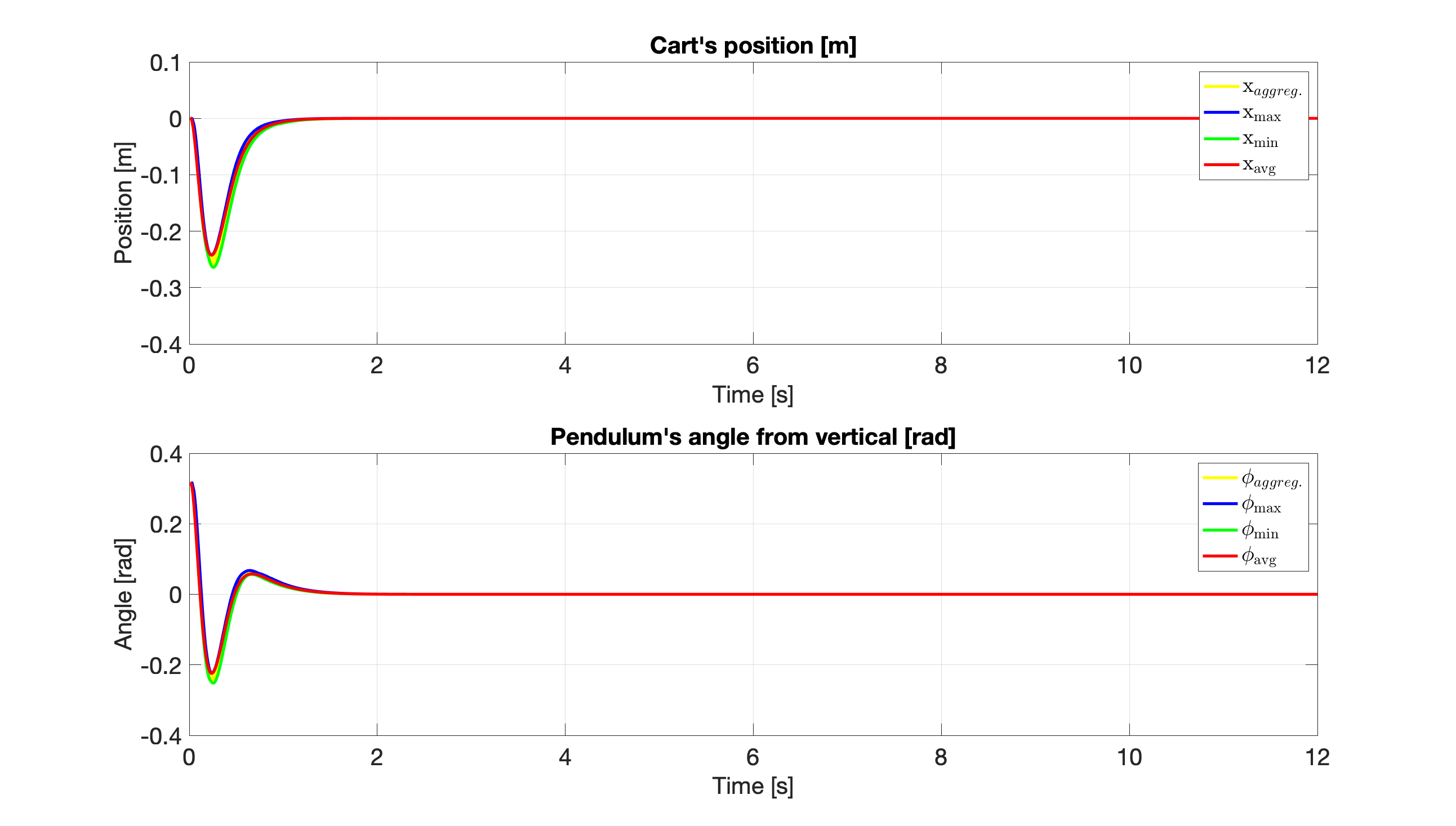}\\
	\vspace{-4mm}	
	\caption{Traces of the system's state that are generated under the Bernoulli control law over WirelessHART channel {\small $\hat{\Gamma}$} having {\small $\hat{d}_{i}\!=\!14\,$}m.}
	\label{fig:bernoulli-14}
\end{figure}

\begin{figure}
	\centering
	\includegraphics[width=0.485\textwidth]{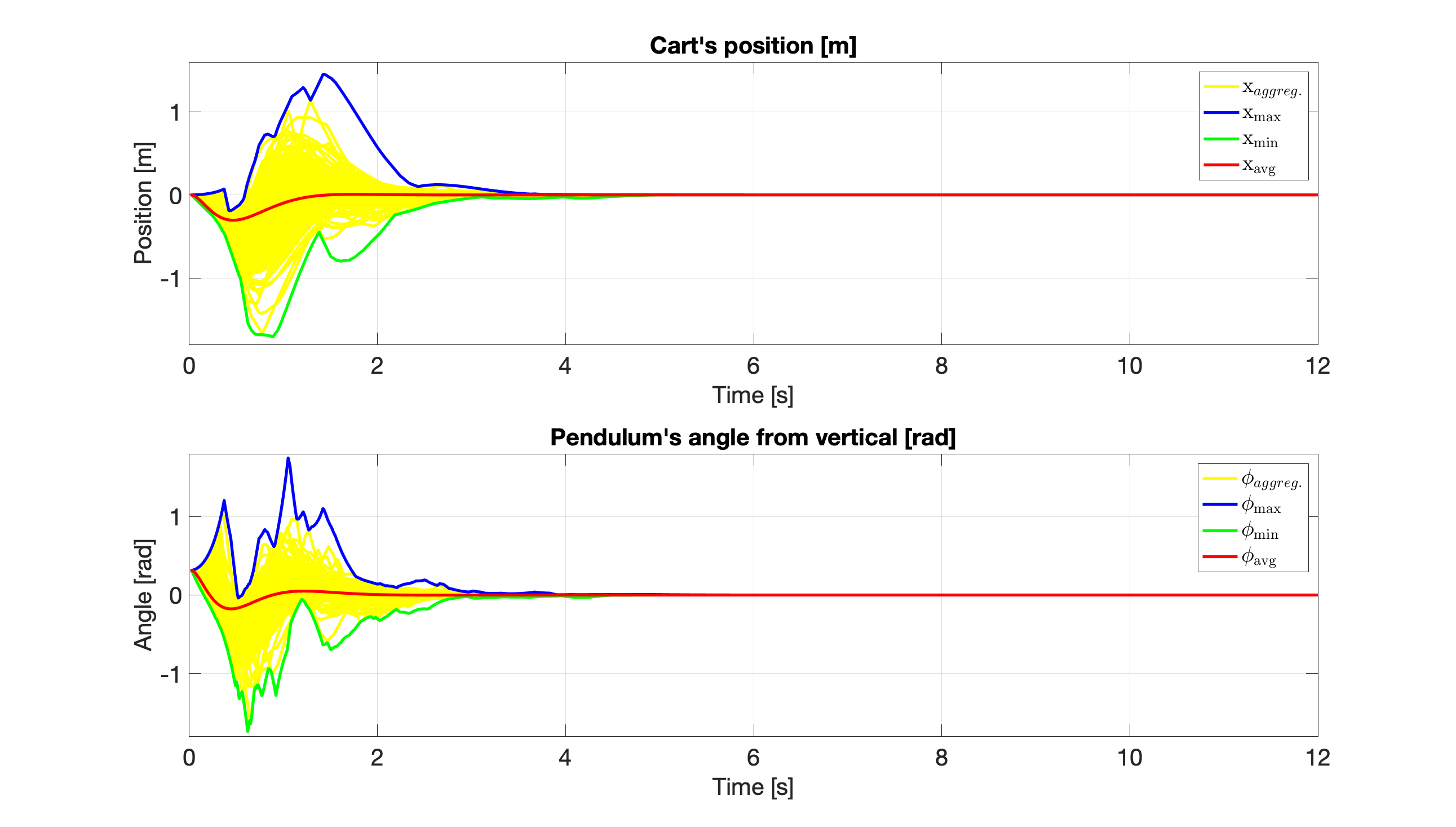}\\
	\vspace{-4mm}	
	\caption{Traces of the system's state that are generated under the Markovian control law over WirelessHART channel {\small $\check{\Gamma}$} having {\small $\check{d}_{i}\!=3.5\!\,$}m.}
	\label{fig:markov-3}
\end{figure}

\begin{figure}
	\centering
	\includegraphics[width=0.485\textwidth]{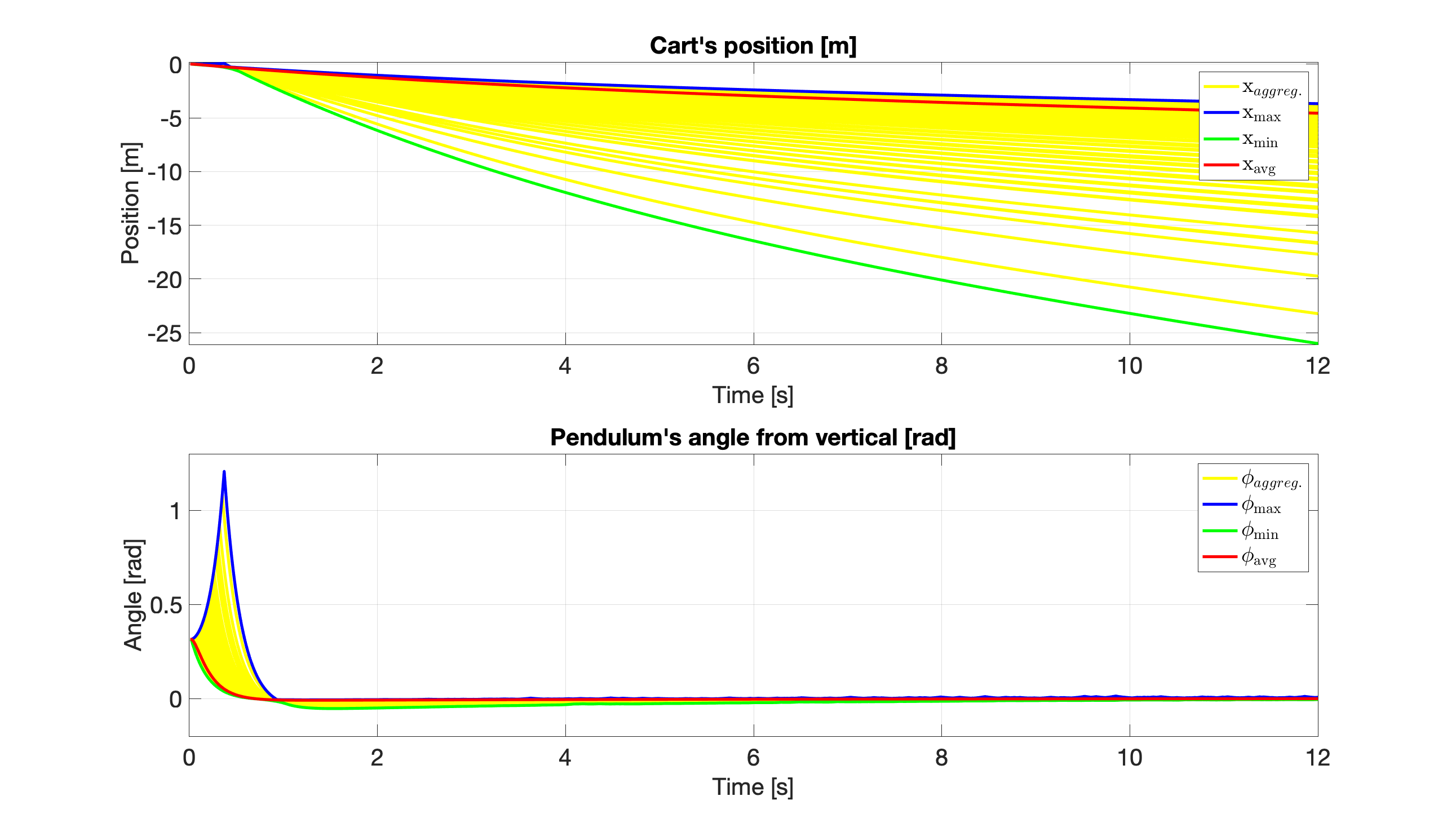}\\
	\vspace{-4mm}	
	\caption{Traces of the system's state that are generated under the Bernoulli control law over WirelessHART channel {\small $\check{\Gamma}$} having {\small $\check{d}_{i}\!=\!3.5\,$}m.}
	\label{fig:bernoulli-3}
\end{figure}

\begin{figure}
	\centering
	\includegraphics[width=0.485\textwidth]{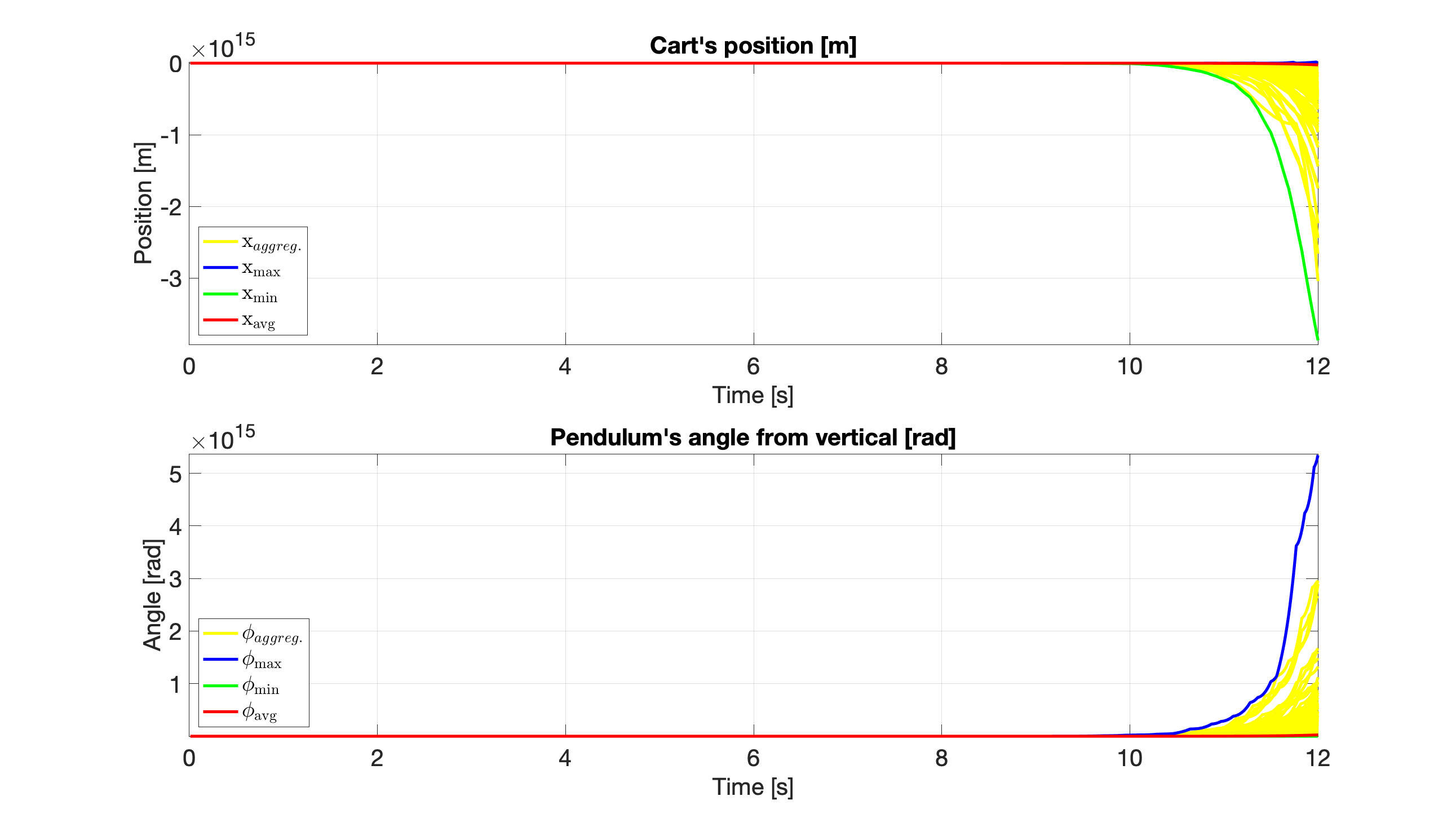}\\
	\vspace{-4mm}	
	\caption{Traces of the system's state that are generated under the Markovian control law over WirelessHART channel {\small $\tilde{\Gamma}$} having {\small $\tilde{d}_{i}\!=2.63\!\,$}m.}
	\label{fig:markov-2}
\end{figure}

\begin{figure}
	\centering
	\includegraphics[width=0.485\textwidth]{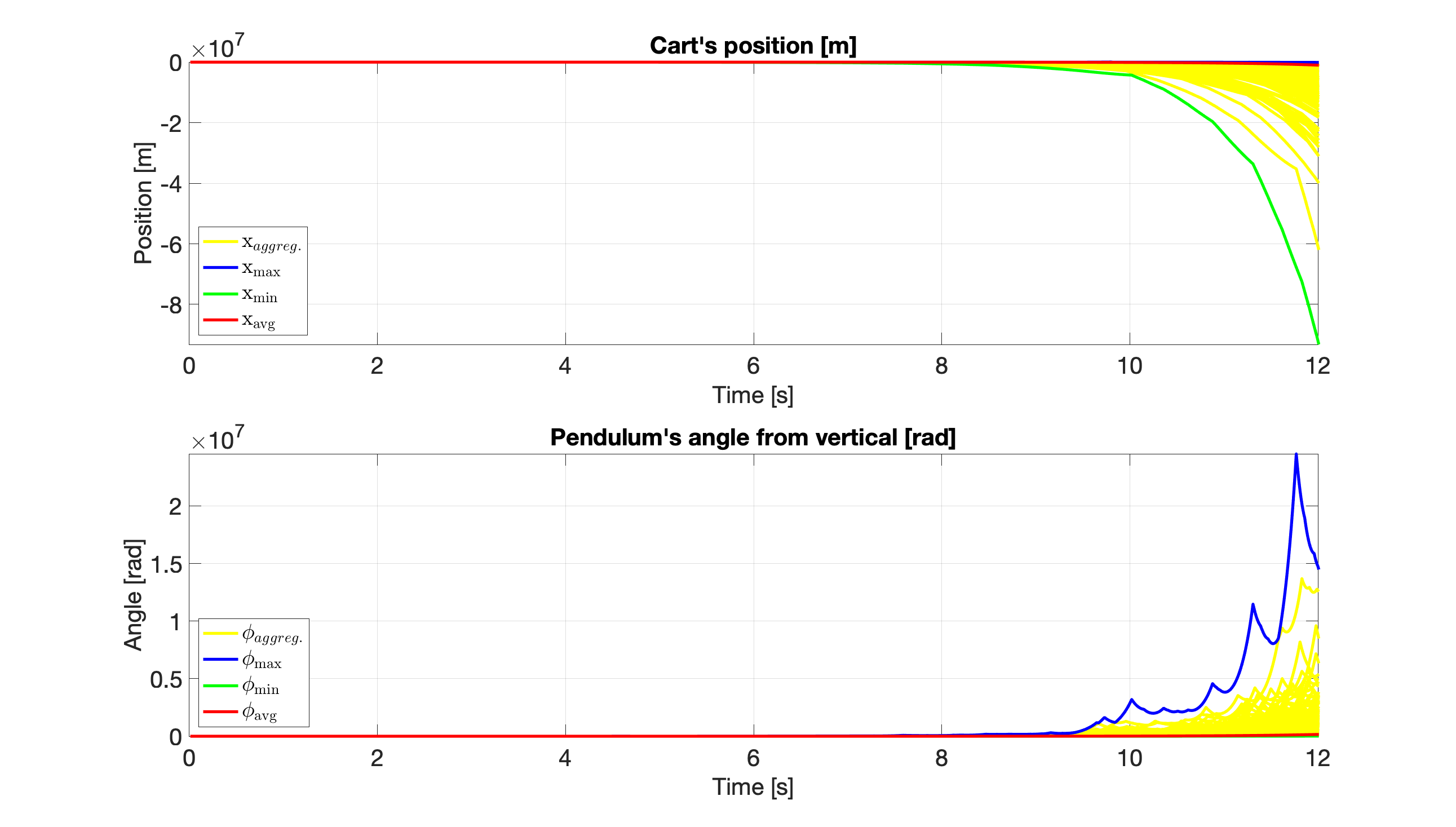}\\
	\vspace{-4mm}	
	\caption{Traces of the system's state that are generated under the Bernoulli control law over WirelessHART channel {\small $\tilde{\Gamma}$} having {\small $\tilde{d}_{i}\!=\!2.63\,$}m.}
	\label{fig:bernoulli-2}
\end{figure}

Figures~\ref{fig:markov-14}\,--\,\ref{fig:bernoulli-2} depict statistical results for simulations of the trajectories generated by
inverted pendulum on a cart, with a remote controller implementing either Bernoulli, or Markovian control, and sending the data over WirelessHART channels {\small $\hat{\Gamma}$}, {\small $\check{\Gamma}$} and {\small $\tilde{\Gamma}$}, respectively. In all cases, 10000 randomly generated admissible evolutions (of length 1200) of the Gilbert channel are emanating from the first mode of operation (i.e. the mode having a certain nonzero probability of packet loss). The same evolutions of a Gilbert channel were used for each pair of Markovian and Bernoulli controllers. 
Since both control strategies do not consider any constraints on the system's states or control inputs, all the physics-related constraints were neglected. It is evident that Figures~\ref{fig:markov-14}, \ref{fig:bernoulli-14}, and \ref{fig:markov-3} show a stable system's behavior, while the behavior illustrated in Figures~\ref{fig:bernoulli-3}, \ref{fig:markov-2}, and \ref{fig:markov-2} is clearly unstable.

%This justifies the effort on providing an accurate channel model, since the simple stationary Bernoulli model 
%cannot provide enough information on stabilizability of the system controlled via a wireless communication 
%protocol such as WirelessHART.

%%%%%%%%%%%%%%%%%%%%%%%%%%%%%%%%%%%%%%%%%%%%%%%%%%%%%%%%%%%%%%%%%%%%%%%%%%%%%%%%
%2345678901234567890123456789012345678901234567890123456789012345678901234567890
%        1         2         3         4         5         6         7         8
%%%%%%%%%%%%%%%%%%%%%%%%%%%%%%%%%%%%%%%%%%%%%%%%%%%%%%%%%%%%%%%%%%%%%%%%%%%%%%%%
% \enlargethispage{-1.5cm} % or
% \addtolength{\textheight}{-14.5cm}
%2345678901234567890123456789012345678901234567890123456789012345678901234567890
%        1         2         3         4         5         6         7         8
%%%%%%%%%%%%%%%%%%%%%%%%%%%%%%%%%%%%%%%%%%%%%%%%%%%%%%%%%%%%%%%%%%%%%%%%%%%%%%%%

%%%%%%%%%%%%%%%%%%%%%%%%%%%%%%%%%%%%%%%%%%%%%%%%%%%%%%%%%%%%%%%%%%%%%%%%%%%%%%%%
%\section*{APPENDIX}
%\section*{ACKNOWLEDGMENTS}
%%%%%%%%%%%%%%%%%%%%%%%%%%%%%%%%%%%%%%%%%%%%%%%%%%%%%%%%%%%%%%%%%%%%%%%%%%%%%%%%
\bibliographystyle{IEEEtran} 
\bibliography{bibliography}

\vspace{1cm} 

\newpage
\section*{APPENDIX}
\normalsize
%%%%%%%%%%%%%%%%%%%%%%%%%%%%%%%%%%%%%%%%%%%%%%%%%%%%%%%%%%%%%%%%%%%%%%%%%%%%%%%%
\subsection{Proof of Proposition \ref{prop:1}}
For \eqref{eq:evolution_qij}, considering \eqref{eq:nu_i},
\eqref{eq:mjls_noiseless}\,--\,\eqref{eq:q}, 
by the low of total probability, one has that

\vspace*{-5mm}
%%%%%%%%%%%%%%%%%%%%%%%%%%%%%%%%%%%%%%%%%%%%%%%%%%%%%%%%%%%%%%%%%%%%%%%%%%%%%%%%
\begin{scriptsize}
\begin{equation*}
\mathrm{m}_{ij}^{(k+1)}\!\!\!=\!\mathbb{E}\!\left(\!x_{k+1}
	\mathbf{1}_{\left\{\theta_{k+1}=j,\,\theta_{k}=i\right\}}\!\right)\!\!=\!\!
 \left(\!\!A\!\!\sum\limits_{\ell=1}^{N}\!\!\!\mathrm{m}_{\ell i}^{(k)}\!\!\!+\!\! 
	B\!\!\sum\limits_{\ell=1}^{N}\!\!K_{\ell} \mathrm{m}_{\ell i}^{(k)} \!\hat{\nu}_i\!\!\right)
	\!p_{ij}.
\end{equation*}
\end{scriptsize}

\vspace*{-4mm}
%%%%%%%%%%%%%%%%%%%%%%%%%%%%%%%%%%%%%%%%%%%%%%%%%%%%%%%%%%%%%%%%%%%%%%%%%%%%%%%%
Similarly, for \eqref{eq:evolution_Qij}, by the low of total probability, 
considering again \eqref{eq:nu_i}, \eqref{eq:mjls_noiseless}\,--\,\eqref{eq:pi_il}, and
\eqref{eq:Q}, one has that

\vspace*{-4mm}
%%%%%%%%%%%%%%%%%%%%%%%%%%%%%%%%%%%%%%%%%%%%%%%%%%%%%%%%%%%%%%%%%%%%%%%%%%%%%%%%
\begin{scriptsize}
\begin{align*}
&\mathrm{M}_{ij}^{(k+1)}\!=\!\mathbb{E}\!\left(x_{k+1}x_{k+1}^{*}
	\mathbf{1}_{\left\{\theta_{k+1}=j,\,\theta_{k}=i\right\}}\right)\!=\nonumber\\[-2mm]
&=\!\!\left(\!\!A\!\!\sum\limits_{\ell=1}^{N}\!\mathrm{M}_{\ell i}^{(k)}\!\!A^{*}\!\!\!+\!\!
	B\!\!\sum\limits_{\ell=1}^{N}\!\!K_{\ell}\mathrm{M}_{\ell i}^{(k)}\!K_{\ell}^{*}\!B^{*}\!
	\hat{\nu}_i\!+\!2\Re\!\!\left(\!\!	A\!\!\sum\limits_{\ell=1}^{N}\!\!
	\mathrm{M}_{\ell i}^{(k)}\!K_{\ell}^{*}\!B^{*}\!\hat{\nu}_i\!\!\right)\!\!\!\right)\!p_{ij},
\end{align*}
\end{scriptsize}

\vspace*{-4mm}
%%%%%%%%%%%%%%%%%%%%%%%%%%%%%%%%%%%%%%%%%%%%%%%%%%%%%%%%%%%%%%%%%%%%%%%%%%%%%%%%
\noindent since from its definition \eqref{eq:Q}, 
{\small $\mathrm{M}_{\ell i}^{(k)}\!\succeq\!0$}. {\small \qed}

%%%%%%%%%%%%%%%%%%%%%%%%%%%%%%%%%%%%%%%%%%%%%%%%%%%%%%%%%%%%%%%%%%%%%%%%%%%%%%%%
\subsection{Proof of Proposition \ref{prop:rho_Lambda_Psi}}
We will follow the line of reasoning of the proof of a similar implication found 
in \cite[Prop. 3.6, p. 35]{costa2006discrete}, which deals with classical MJLSs, 
having the operation modes observable instantaneously.
Let {\small $\left\{\tilde{e}_i\right\}_{i=1}^{n_{x}}$}, 
{\small $\left\{\hat{e}_i\right\}_{i=1}^{N^{2}n_{x}}$} and 
{\small $\left\{\check{e}_i\right\}_{i=1}^{N^{2}n_{x}^{2}}$}
be the canonical orthonormal basis for the linear spaces
{\small $\mathbb{F}^{n_{x}}$}, {\small $\mathbb{F}^{N^{2}n_{x}}$} and 
{\small $\mathbb{F}^{N^{2}n_{x}^{2}}$}, respectively. Fix arbitrarily 
{\small $\zeta\!\in\!\left\{i\right\}_{i=1}^{n_x}$}, 
{\small $l\!\in\!\left\{i\right\}_{i=1}^{N}$}, and
{\small $\iota\!\in\!\left\{i\right\}_{i=1}^{N}$}. 
Consider the system \eqref{eq:mjls_noiseless}, with the initial conditions
{\small $\theta_{-1}\!=\!l$, $\theta_{0}\!=\!\iota$}, and {\small $x_0\!=\!\tilde{e}_{\zeta}$}.
Then, the only element {\small $\mathrm{m}_{\ell i}^{(0)}$} different 
from the vector of all zeros is {\small $\mathrm{m}_{l\iota}^{(0)}\!=\!\tilde{e}_{\zeta}$}, 
and $\mathrm{vec}^{2}\!\left({\mathbf{m}}^{(0)}\right)\!=\!\hat{e}_{\kappa}$, where
{\small $\kappa\!=\!\zeta\!+\!n_{x}\!\left(l\!-\!1\!+\!N\left(\iota\!-\!1\right)\!\right)$}.
Similarly, one has that the only element {\small $\mathrm{M}_{\ell i}^{(0)}$} different from the null matrix is 
{\small $\mathrm{M}_{l\iota}^{(0)}\!=\!\tilde{e}_{\zeta}\tilde{e}_{\zeta}^{*}$}, and
{\small $\mathrm{vec}^{2}\!\left(\mathbf{M}^{(0)}\right)\!=\!\check{e}_{\varkappa}$}, where
{\small $\varkappa\!=\!\zeta\!+\!n_{x}\!\left(\zeta\!-\!1\right)\!+\!
n_{x}^{2}\!\left(l\!-\!1\!+\!N\left(\iota\!-\!1\right)\!\right)$}.
Now, on one hand one has from the repeated application of \eqref{eq:vec2q} that {\small $\|\mathbf{m}^{(k)}\|_{\bm 2}^{2} = \left\|{\bm \Psi}^{k}\hat{e}_{\kappa}\right\|_{2}^{2}$}.
%%%%%%%%%%%%%%%%%%%%%%%%%%%%%%%%%%%%%%%%%%%%%%%%%%%%%%%%%%%%%%%%%%%%%%%%%%%%%%%%
On the other hand, from \eqref{eq:q}, by the triangle inequality, linearity of the 
expected value and of the trace, and by the definition of the trace, one has that

\vspace*{-4mm}
%%%%%%%%%%%%%%%%%%%%%%%%%%%%%%%%%%%%%%%%%%%%%%%%%%%%%%%%%%%%%%%%%%%%%%%%%%%%%%%%
\begin{scriptsize}
\begin{equation*}
\left\|\mathbf{m}^{(k)} \right\|_{\bm 2}^{2} \!\!=\!\!
\sum\nolimits_{\ell=1}^{N}\!\sum\nolimits_{i=1}^{N}\!\left\|\mathrm{m}_{\ell i}^{(k)}\right\|_{2}^{2} \!\leq\! 
\sum\nolimits_{\ell=1}^{N}\!\sum\nolimits_{i=1}^{N}\left\|\mathrm{M}_{\ell i}^{(k)}\right\|_{\bm 1}\!=\! \left\|{\bm \Lambda}^{\!k}\check{e}_{\varkappa}\right\|_{1}.
\end{equation*}
\end{scriptsize}

\vspace*{-5mm}
%%%%%%%%%%%%%%%%%%%%%%%%%%%%%%%%%%%%%%%%%%%%%%%%%%%%%%%%%%%%%%%%%%%%%%%%%%%%%%%%
Now, the hypothesis {\small $\rho\!\left({\bm \Lambda}\right)\!<\!1\!\Rightarrow\!
\lim\nolimits_{k\to\infty}{\bm \Lambda}^{\!k}\!=\!0$}. Thus, 
{\small $\lim\nolimits_{k\to\infty}\left\|{\bm \Psi}^{k}\hat{e}_{\kappa}\right\|_{2}^{2}
\!=\!0$}. Since {\small $\zeta$}, {\small $l$} and {\small $\iota$} were chosen 
arbitrarily, it follows that the previous expression holds 
{\small $\forall \kappa\!\in\!\left\{i\right\}_{i=1}^{N^{2}n_{x}}$}.
Hence, {\small $\lim\nolimits_{k\to\infty}\left\|{\bm \Psi}^{k}{\bm v}\right\|_{2}^{2}\!=\!0$},
{\small $\forall {\bm v}\!\in\!\mathbb{F}^{N^{2}n_{x}}$}, which implies that
{\small $\lim\nolimits_{k\to\infty}{\bm \Psi}^{k}\!=\!0$}, and proves the thesis, 
{\small $\rho\!\left({\bm \Psi}\right)\!<\!1$}. \qed
%%%%%%%%%%%%%%%%%%%%%%%%%%%%%%%%%%%%%%%%%%%%%%%%%%%%%%%%%%%%%%%%%%%%%%%%%%%%%%%%%

\subsection{Proof of Proposition \ref{prop:delayedNoiselessMeanSquareStability}}
For finite-dimensional linear spaces all norms are equivalent
\cite[Theorem 4.27]{kubrusly2001elements}, so one can choose any particular norm 
in \eqref{eq:mss_general}. We first prove that 
{\small $\rho\!\left({\bm \Lambda}\right)\!<\!1$} implies the mean square stability of 
the system \eqref{eq:mjls_noiseless}. 
From the definition of the {\small $\ell_1$} norm, triangle inequality and 
\eqref{eq:moments}, one finds that

\vspace*{-4mm}
%%%%%%%%%%%%%%%%%%%%%%%%%%%%%%%%%%%%%%%%%%%%%%%%%%%%%%%%%%%%%%%%%%%%%%%%%%%%%%%%
\begin{scriptsize}
\begin{equation*}
\left\|\mathbf{M}^{(k)} \right\|_{\bm 1} \!\!\!=\!\!
\sum\nolimits_{\ell=1}^{N}\!\sum\nolimits_{i=1}^{N}\!
\left\|\mathrm{M}_{\ell i}^{(k)} \right\|_{\bm 1} \!\!\geq\!
\left\|\sum\nolimits_{\ell=1}^{N}\!\sum\nolimits_{i=1}^{N}\!\!\mathrm{M}_{\ell i}^{(k)}\right\|_{\bm 1} 
\!\!\!=\!\left\|\mathbb{E}\!\left(x_{k}x_{k}^{*}\right)\right\|_{\bm 1}\!\!.
\end{equation*}
\end{scriptsize}

\vspace*{-6mm}
%%%%%%%%%%%%%%%%%%%%%%%%%%%%%%%%%%%%%%%%%%%%%%%%%%%%%%%%%%%%%%%%%%%%%%%%%%%%%%%%
By induction from \eqref{eq:vec2Q} and the definition of the {\small $\ell_1$} 
norm, 
% % one has that 
{\small $\|\mathbf{M}^{(k)} \|_{\bm 1} =
\|\mathrm{vec}^{2}\!(\!\mathbf{M}^{(k)}\!)\|_{1} =
\|{\bm \Lambda}^{k}\mathrm{vec}^{2}\!(\!\mathbf{M}^{(0)}\!)\|_{1}$}.
%%%%%%%%%%%%%%%%%%%%%%%%%%%%%%%%%%%%%%%%%%%%%%%%%%%%%%%%%%%%%%%%%%%%%%%%%%%%%%%%
Since {\small $\rho\!\left({\bm \Lambda}\right)\!<\!1$} implies that
{\small $\lim\nolimits_{k\to\infty}{\bm \Lambda}^{\!k}\!=\!0$}, so that 
{\small $\lim\nolimits_{k\to\infty}\|\mathbf{M}^{(k)} \|_{\bm 1}\!\!=\!0$},
and, consequently, 
{\small $\lim\nolimits_{k\to\infty}\left\|
\mathbb{E}\left(x_{k}x_{k}^{*}\right)\!-\!0\right\|_{\bm 1}\!\!=\!0$}.

%%%%%%%%%%%%%%%%%%%%%%%%%%%%%%%%%%%%%%%%%%%%%%%%%%%%%%%%%%%%%%%%%%%%%%%%%%%%%%%%
By following exactly the same line of reasoning, from 
Proposition \ref{prop:rho_Lambda_Psi}, the definition of the 
{\small $\ell_1$} norm, triangle inequality, \eqref{eq:moments}, and the repeated 
application of \eqref{eq:vec2q}, one finds that 
{\small $\|\mathbf{m}^{(k)} \|_{\bm 1} \!=\! \|\mathbb{E}(x_{k})\|_{1}$}, 
{\small $\|\mathbf{m}^{(k)} \|_{\bm 1} \!=\! \|{\bm \Psi}^{k}\mathrm{vec}^{2}(\!\mathbf{m}^{(0)}\!)\|_{1}$} 
and 
{\small $\rho({\bm\Lambda})\!<\!1\!\Rightarrow\! \lim\nolimits_{k\to\infty}\|\mathbb{E}(x_{k})\!-\!0\|_{1}\!\!=\!0$}
%%%%%%%%%%%%%%%%%%%%%%%%%%%%%%%%%%%%%%%%%%%%%%%%%%%%%%%%%%%%%%%%%%%%%%%%%%%%%%%%
{\small $\forall x_{k}$}, and the first part of the proof is concluded.

So, it remains to prove the sufficiency, i.e., the conditions in \eqref{eq:mss_general}
imply that {\small $\rho\!\left({\bm\Lambda}\right)\!<\!1$}. 
By hypothesis, one has that
{\small $\lim\nolimits_{k\to\infty}\left\|\mathbb{E}\!\left(x_{k}x_{k}^{*}\right)
\!-\!\mathrm{M}_{e}\right\|\!=\!0$} for all
initial condition {\small $\left(\hat{x}_{0},\theta_{0}\right)$}. By taking 
{\small $\hat{x}_{0}\!=\!0$}, one finds that
{\small $\mathrm{M}_{e}$} must necessarily be equal to zero, and 
{\small $\lim\nolimits_{k\to\infty}\mathbb{E}\!\left(x_{k}x_{k}^{*}\right)\!=\!0$}.
Then, from \eqref{eq:moments}, it follows that
{\small $\lim\nolimits_{k\to\infty}\sum\nolimits_{\ell=1}^{N}\sum\nolimits_{i=1}^{N}
\mathrm{M}_{\ell i}^{(k)}\!=\!0$}, with 
{\small $\mathrm{M}_{\ell i}^{(k)}\!\succeq\!0$}, {\small $\forall k,\ell,i$}. Thus, from
\eqref{eq:Q}, one has that {\small $\lim\nolimits_{k\to\infty}\mathbf{M}^{(k)}\!=\!0$}. 
Since the linear mapping {\small $\mathrm{vec}^{2}\!\left(\cdot\right)$} is 
uniform homeomorphic (see e.g. \cite{naylor2000linear} for additional details), 
the convergent behaviour of {\small $\mathbf{M}^{(k)}$} is preserved by 
{\small $\mathrm{vec}^{2}\!\left(\!\mathbf{M}^{(k)}\!\right)$}. So, from the
repeated application of \eqref{eq:vec2Q}, one obtains that
{\small $\lim\nolimits_{k\to\infty}{\bm\Lambda}^{k}\mathrm{vec}^{2}\!
\left(\!\mathbf{M}^{(0)}\!\right)\!=\!0$}. This last statement is true
{\small $\forall\mathbf{M}^{(0)}$} if and only if (from now on, iff)
{\small $\lim\nolimits_{k\to\infty}{\bm\Lambda}^{k}\!=\!0$}, i.e., iff
{\small $\rho\!\left({\bm\Lambda}\right)\!<\!1$}. \qed

%%%%%%%%%%%%%%%%%%%%%%%%%%%%%%%%%%%%%%%%%%%%%%%%%%%%%%%%%%%%%%%%%%%%%%%%%%%%%%%%
\subsection{Proof of Proposition \ref{prop:delayedNoiselessLyapunov}}
The proof of necessity follows the same steps of the proof of 
\cite[Theorem~3.19, p.~41]{costa2006discrete}, if one considers the system described
by the following recursive equation {\small $\mathbf{T}^{(k+1)} \!=\! \mathcal{T}\!(\!\mathbf{T}^{(k)}\!)\!\!\;$, $\mathbf{T}^{(0)}\!\!\in\!\mathbb{F}^{Nn_x\times Nn_x}$, and $~\mathbf{T}^{(0)}\!\succ\!0$}.
An interested reader may also refer to \cite[Lemmas 47,~54, pp.~214\,--\,216]{callier1991linear} 
for additional details on the proof for the simple linear case (without jumps).
The proof of sufficiency instead is obtained by following the line of reasoning of
the aforementioned \cite[Lemma 54, pp.~215\,--\,216]{callier1991linear}, and 
\cite[Lemma~1]{kubrusly1985mean}. \qed

%%%%%%%%%%%%%%%%%%%%%%%%%%%%%%%%%%%%%%%%%%%%%%%%%%%%%%%%%%%%%%%%%%%%%%%%%%%%%%%%
\subsection{Proof of Proposition \ref{prop:noisyQ}}
By hypothesis on the process noise, {\small $\forall k\!\in\!\mathbb{N}_{0}$}, 
one has that {\small $\mathbb{E}\!\left(w_{k}\right)\!=\!0$} and 
{\small $\mathbb{E}\!\left(w_{k}w_{k}^{*}\right)\!=\!\Sigma_{w}$}, and
the first statement is obtained by the same line of reasoning of the first part
of the proof of Proposition \ref{prop:1}, after taking into account that 
{\small $w_k$} is independent from {\small $\theta_{k}$} and {\small $\theta_{k-1}$}. 

%%%%%%%%%%%%%%%%%%%%%%%%%%%%%%%%%%%%%%%%%%%%%%%%%%%%%%%%%%%%%%%%%%%%%%%%%%%%%%%%
The expression \eqref{eq:evolution_noisy_Qij} is derived in similar fashion
from the law of total probability, the aforementioned hypothesis, 
\eqref{eq:nu_i}, \eqref{eq:p_lij_0}\,--\,\eqref{eq:pi_il}, 
\eqref{eq:Q}, and \eqref{eq:mjls_alternative}, observing that 
{\small $\mathrm{M}_{\ell i}^{(k)}\!\succeq\!0$}. \qed

%%%%%%%%%%%%%%%%%%%%%%%%%%%%%%%%%%%%%%%%%%%%%%%%%%%%%%%%%%%%%%%%%%%%%%%%%%%%%%%%
\subsection{Proof of Theorem \ref{theorem:delayedMeanSquareStability}}
We prove first that {\small $\rho\!\left({\bm \Lambda}\right)\!<\!1$} implies 
that under the stated assumptions the system \eqref{eq:mjls_alternative} is mean 
square stable. 

%%%%%%%%%%%%%%%%%%%%%%%%%%%%%%%%%%%%%%%%%%%%%%%%%%%%%%%%%%%%%%%%%%%%%%%%%%%%%%%%
Let {\small $\mathbb{E}\!\left(\mathbf{1}_{\left\{\theta_{k}=i\right\}}\right)\!
\triangleq\!\chi_{i}^{(k)}$} be the probability mass function of the MC
$\Theta$. By the definition of the conditional probability,
{\small $\pi_{\ell i}^{(k)}\!=\! p_{\ell i}\chi_{\ell}^{(k-1)}$}. So,
{\small $\chi_{i}^{(k)} \!=\! \sum\nolimits_{\ell=1}^{N} \pi_{\ell i}^{(k)}$}. 
By hypothesis, $\Theta$ is ergodic, i.e., for any given initial probability 
distribution {\small $\left\{\hat{\chi}_{i}^{(0)}\right\}_{i=1}^{\!N}$}, there 
exists a limit probability distribution 
{\small $\left\{\hat{\chi}_{i}^{(\infty)}\right\}_{i=1}^{\!N}$} which does not 
depend on {\small $\left\{\hat{\chi}_{i}^{(0)}\right\}_{i=1}^{\!N}$}, such that 
{\small $\sum\nolimits_{i=1}^{N} p_{ij}\hat{\chi}_{i}^{(\infty)}\!=\!\hat{\chi}_{j}^{(\infty)}$, 
$\sum\nolimits_{i=1}^{N} \hat{\chi}_{i}^{(\infty)}\!=\!1$ and 
$|\chi_{i}^{(k)}\!-\hat{\chi}_{i}^{(\infty)} |\!\leq\!\eta\varepsilon^{k}$},
%%%%%%%%%%%%%%%%%%%%%%%%%%%%%%%%%%%%%%%%%%%%%%%%%%%%%%%%%%%%%%%%%%%%%%%%%%%%%%%%
for some {\small $\eta\!\geq\!0$} and {\small $0\!<\!\varepsilon\!<\!1$} 
(cf. \cite[p. 48]{costa2006discrete}).
%%%%%%%%%%%%%%%%%%%%%%%%%%%%%%%%%%%%%%%%%%%%%%%%%%%%%%%%%%%%%%%%%%%%%%%%%%%%%%%%
Let {\small 
$\hat{\pi}_{\ell i}^{(\infty)}\!\!=\! p_{\ell i}\hat{\chi}_{\ell}^{(\infty)}\!$}, %%%and
{\small $\hat{\bm \Pi}^{(\infty)}\!\!=\!\!\Big[\hat{\pi}_{\ell i}^{(\infty)}\!\!
\otimes\!\mathbb{I}_{n_x}\Big]_{\ell,i=1}^{N}\!$}.
Then, from the ergodic assumption, %%% one has that
{\small $\sum\nolimits_{\ell=1}^{N}\hat{\pi}_{\ell i}^{(\infty)}\!=\!
\hat{\chi}_{i}^{(\infty)}$}, {\small $\sum\nolimits_{\ell=1}^{N}
\sum\nolimits_{i=1}^{N}\hat{\pi}_{\ell i}^{(\infty)}\!=\!1$}, and, 
since by its definition {\small $p_{\ell i}\!\geq\!0$}, we have that
{\small $p_{\ell i}\left|\chi_{\ell}^{(k)}\!-\hat{\chi}_{\ell}^{(\infty)}\right|
\!\leq\!p_{\ell i}\eta\varepsilon^{k}$}, %%%and it follows that

\vspace*{-1mm}
\begin{small}
\begin{equation}\label{eq:limit_pi_bound}
\left|\pi_{\ell i}^{(k+1)}\!-\hat{\pi}_{\ell i}^{(\infty)}\right|\!\leq\!
p_{\ell i}\eta\varepsilon^{k}, ~ \eta\!\geq\!0, ~ 0\!<\!\varepsilon\!<\!1.
\end{equation}
\end{small}
\vspace*{-5mm}

%%%%%%%%%%%%%%%%%%%%%%%%%%%%%%%%%%%%%%%%%%%%%%%%%%%%%%%%%%%%%%%%%%%%%%%%%%%%%%%%
We know from Proposition \ref{prop:noisyQ} that the first and the
second moments of the system's state evolve according to \eqref{eq:vec2q} and 
\eqref{eq:vec2S_noisy}, respectively. By Proposition \ref{prop:rho_Lambda_Psi}, 
{\small $\rho\!\left({\bm \Lambda}\right)\!<\!1\!\Rightarrow\!
\rho\!\left({\bm \Psi}\right)\!<\!1$}. Thus, 
{\small $\lim\limits_{k\to\infty}\!{\bm \Psi}^{k}\!\!=\!0$}, and
{\small $\lim\limits_{k\to\infty}\left\|\mathbb{E}\left(x_{k}\right)\!-\!x_e
\right\|\!=\!0$}, where {\small $x_e\!=\!0$}, for all initial conditions 
{\small $\left(\hat{x}_{0},\theta_{0}\right)$}. Now, in order to show that also 
{\small $\lim\limits_{k\to\infty}\left\|\mathbb{E}\left(x_{k}x_{k}^{*}\right)\!-\!
{M}_{e}\right\|\!=\!0$}, we will prove that 
{\small $\left(\!\mathcal{Z}_k\right)_{k=0}^{\infty}$}, with 
{\small $\mathcal{Z}_k\!=\!{\bm \Upsilon}\mathrm{vec}^{2}\!\left(\!{\bm \Pi}^{(k)}
\!\right)\!$}, is a Cauchy summable sequence, i.e., it is a Cauchy sequence in a 
complete normed space {\small $\mathbb{F}^{N^{2} n_{x}^{2}}$}, and 
{\small $\sum\limits_{k=0}^{\infty} \sup\limits_{\tau\geq 0}\left\|
\mathcal{Z}_{k+\tau}\!-\!\mathcal{Z}_k\right\|\!<\!\infty$}, so that 
{\small $\mathrm{vec}^2\!\!\left(\!\mathbf{M}^{(k)}\!\right)$} from 
\eqref{eq:vec2S_noisy} is also Cauchy summable, and for any initial condition 
{\small $\mathrm{vec}^2\!\!\left(\!\mathbf{M}^{(0)}\!\right)$}, by 
\cite[Proposition 2.9, p. 20]{costa2006discrete} 
{\small $\lim\limits_{k\to\infty}\mathrm{vec}^2\!\!\left(\!\mathbf{M}^{(k)}
\!\right)\!=\!\left(\mathbb{I}_{N^{2}n_{x}^{2}}\!-\!{\bm \Lambda}\right)^{\!-1}
\!{\bm \Upsilon}\!\lim\limits_{k\to\infty}\mathrm{vec}^{2}\!
\left(\!{\bm \Pi}^{(k)}\!\right)$}.
%%%%%%%%%%%%%%%%%%%%%%%%%%%%%%%%%%%%%%%%%%%%%%%%%%%%%%%%%%%%%%%%%%%%%%%%%%%%%%%%
Since for a finite-dimensional linear spaces all norms are equivalent, 
we will use {\small $\ell_1$} norm to prove first that the elements of the sequence {\small $\left(\mathcal{Z}_k\right)_{k=0}^{\infty}$} become arbitrarily close to each other as the sequence progresses, i.e., {\small $\left(\mathcal{Z}_k\right)_{k=0}^{\infty}$} is a Cauchy sequence. Formally, {\small $\forall t, k$}, from the definition of the {\small $\ell_1$}-norm, triangle inequality, 
additivity of the linear mapping {\small $\mathrm{vec}^{2}\!\left(\cdot\right)$}, and \eqref{eq:limit_pi_bound}

\vspace*{-2mm}
%%%%%%%%%%%%%%%%%%%%%%%%%%%%%%%%%%%%%%%%%%%%%%%%%%%%%%%%%%%%%%%%%%%%%%%%%%%%%%%%
\begin{scriptsize}
\begin{align}\label{eq:z_cauchy}
&\left\|\mathcal{Z}_{t}-\!\mathcal{Z}_{k}\right\|_{1}\!=\!
%\Big\|{\bm \Upsilon}\!
%\Big(\!\mathrm{vec}^{2}\!\Big(\!{\bm \Pi}^{(t)}\!\!-\!\hat{\bm \Pi}^{(\infty)}
%\!\!+\!\hat{\bm \Pi}^{(\infty)}\!\!-\!{\bm \Pi}^{(k)}\!\Big)\!\Big)\!\Big\|_{1}
%\!=\! \nonumber \\
\Big\|{\bm \Upsilon}\!\Big(\!\mathrm{vec}^{2}\!\Big({\bm \Pi}^{(t)}\!\!-\!
\hat{\bm \Pi}^{(\infty)}\!\Big)\!+\mathrm{vec}^{2}\!\Big(\hat{\bm\Pi}^{(\infty)}\!
\!-\!{\bm \Pi}^{(k)}\!\Big)\!\!\Big)\!\Big\|_{1} \!\leq\!  \\
&\big\|{\bm \Upsilon}\big\|_{\bm 1}\!\bigg(\!\Big\|{\bm \Pi}^{(t)}\!\!-\!
\hat{\bm \Pi}^{(\infty)}\Big\|_{\bm 1}\!\!\!+\!\Big\|\hat{\bm \Pi}^{(\infty)}
\!\!-\!{\bm \Pi}^{(k)}\Big\|_{\bm 1}\!\bigg) \!\leq\!
N^{2}n_{x}\big\|{\bm \Upsilon}\big\|_{\bm 1}\!\eta\!\left(
\varepsilon^{t-1}\!\!+\!\varepsilon^{k-1}\right)\!, \nonumber
\end{align}
\end{scriptsize}

\vspace*{-4mm}
%%%%%%%%%%%%%%%%%%%%%%%%%%%%%%%%%%%%%%%%%%%%%%%%%%%%%%%%%%%%%%%%%%%%%%%%%%%%%%%%
\noindent for some {\small $\eta\!\geq\!0$}, {\small $0<\!\varepsilon\!<\!1$}, 
proving that {\small $\left(\mathcal{Z}_k\right)_{k=0}^{\infty}$} is a 
Cauchy sequence, with {\small $\!\lim\limits_{k\to\infty}\!\!\mathcal{Z}_{k}\!=\!
{\bm \Upsilon}\mathrm{vec}\Big(\!\hat{\bm \Pi}^{(\infty)}\!\Big)\!$}. Also, for {\small $t\!=\!k\!+\!\tau$}, 
\eqref{eq:z_cauchy} implies that	 {\small $\sum\limits_{k=0}^{\infty}\sup\limits_{\tau\geq 0}\|
\mathcal{Z}_{k+\tau}\!-\!\mathcal{Z}_k\|_{1}\!\leq\! \frac{2N^{2}n_{x}\eta}{\varepsilon(1\!-\!\varepsilon)}
\big\|{\bm \Upsilon}\big\|_{\bm 1} \!<\!\infty$}.
%%%%%%%%%%%%%%%%%%%%%%%%%%%%%%%%%%%%%%%%%%%%%%%%%%%%%%%%%%%%%%%%%%%%%%%%%%%%%%%%
where the last equality is obtained from the formula of the sum of a 
geometric series. This proves that the sequence 
{\small $\left(\mathcal{Z}_k\right)_{k=0}^{\infty}$} is Cauchy summable. Thus,
by \cite[Proposition 2.9, p. 20]{costa2006discrete}, if 
{\small $\rho\!\left({\bm \Lambda}\right)\!<\!1$}, then also 
{\small $\mathrm{vec}^2\!\!\left(\!\mathbf{M}^{(k)}\!\right)$} from \eqref{eq:vec2S_noisy}
is a Cauchy summable sequence, and for any initial condition 
{\small $\mathrm{vec}^2(\mathbf{M}^{(0)})$, $\lim\limits_{k\to\infty}\mathrm{vec}^2(\mathbf{M}^{(k)}
)=(\mathbb{I}_{N^{2}n_{x}^{2}}-{\bm \Lambda})^{-1}
{\bm \Upsilon}\mathrm{vec}^{2}\!\Big(\hat{\bm \Pi}^{(\infty)}\Big)$}.
%%%%%%%%%%%%%%%%%%%%%%%%%%%%%%%%%%%%%%%%%%%%%%%%%%%%%%%%%%%%%%%%%%%%%%%%%%%%%%%%
Since {\small $\mathrm{vec}^{2}\!\left(\cdot\right)$} is uniform homeomorphic,
it follows that {\small $\lim\limits_{k\to\infty}\!\mathbf{M}^{(k)} = \mathbf{M}^{(\infty)}$}.
%%%%%%%%%%%%%%%%%%%%%%%%%%%%%%%%%%%%%%%%%%%%%%%%%%%%%%%%%%%%%%%%%%%%%%%%%%%%%%%%
Thus, together with \eqref{eq:Q} and \eqref{eq:moments}, it implies that {\small $\lim_{k \to \infty}\|\mathbb{E}(x_k x_k^* )-\sum\limits_{\ell=1}^{N}\sum\limits_{i=1}^{N} \mathrm{M}_{\ell i}^{(\infty)}\|=0$}.
%%%%%%%%%%%%%%%%%%%%%%%%%%%%%%%%%%%%%%%%%%%%%%%%%%%%%%%%%%%%%%%%%%%%%%%%%%%%%%%%
Hence, the system \eqref{eq:mjls_alternative} is mean square stable.
%%%%%%%%%%%%%%%%%%%%%%%%%%%%%%%%%%%%%%%%%%%%%%%%%%%%%%%%%%%%%%%%%%%%%%%%%%%%%%%%
So, it remains to prove the necessity, i.e., if the system 
\eqref{eq:mjls_alternative} is mean square stable, 
then {\small $\rho\!\left({\bm \Lambda}\right)\!<\!1$}. 
From \eqref{eq:joint_probability}, it is immediate to verify that

\vspace*{-4mm}
%%%%%%%%%%%%%%%%%%%%%%%%%%%%%%%%%%%%%%%%%%%%%%%%%%%%%%%%%%%%%%%%%%%%%%%%%%%%%%%%
\begin{small}
\begin{equation}\label{eq:evolutionPi}
\mathrm{vec}^{2}\!\!\left(\!{\bm\Pi}^{(k)}\!\right)\!\!=\!
{\bm\Xi}\,\mathrm{vec}^{2}\!\!\left(\!{\bm\Pi}^{(k-1)}\!\right)\!,\quad
{\bm\Xi}\!=\!\Delta\mathrm{P}_{1}\!\otimes\!\!\left(\bigconcatenate\limits_{j=1}^{N}
	\mathbb{I}_{n_{x}^{2}}\!\right)\!,
\end{equation}
\end{small} 

\vspace*{-4mm}
%%%%%%%%%%%%%%%%%%%%%%%%%%%%%%%%%%%%%%%%%%%%%%%%%%%%%%%%%%%%%%%%%%%%%%%%%%%%%%%%
\noindent and from \eqref{eq:vec2S_noisy} we have that

\vspace*{-3mm}
%%%%%%%%%%%%%%%%%%%%%%%%%%%%%%%%%%%%%%%%%%%%%%%%%%%%%%%%%%%%%%%%%%%%%%%%%%%%%%%%
\begin{scriptsize}
\begin{equation}\label{eq:vec2S_noisy_explicit}
\mathrm{vec}^{2}\!\!\left(\!\mathbf{M}^{(k+1)}\!\right)\!=\!
	\bm{\Lambda}^{\!k}\mathrm{vec}^{2}\!\!\left(\!\mathbf{M}^{(0)}\!\right) \!+\!
	\!\sum\nolimits_{t=0}^{k-1}\!{\bm\Lambda}^{\!t}{\bm\Upsilon}{\bm\Xi}^{k-t-1}
	\mathrm{vec}^{2}\!\!\left(\!{\bm \Pi}^{(0)}\!\right)\!,
\end{equation}
\end{scriptsize} 

\vspace*{-5mm}
%%%%%%%%%%%%%%%%%%%%%%%%%%%%%%%%%%%%%%%%%%%%%%%%%%%%%%%%%%%%%%%%%%%%%%%%%%%%%%%%
\noindent where, from \eqref{eq:Q}, \eqref{eq:Lambda}, \eqref{eq:Upsilon} and 
\eqref{eq:evolutionPi}, only the first addend depends on the initial state $x_0$. 
By hypothesis the system 
is mean square stable, so,
from \eqref{eq:mss_general} 
and \eqref{eq:moments},
there exists {\small $M_e$} (depending only on the process noise characteristics) such 
that {\small $\lim\limits_{k\to\infty}\!\mathbb{E}\!\left(x_{k}x_{k}^{*}\right)\!=\!M_e$} for any 
%%%%%%%%%%%%%%%%%%%%%%%%%%%%%%%%%%%%%%%%%%%%%%%%%%%%%%%%%%%%%%%%%%%%%%%%%%%%%%%%
{\small $\mathbb{E}\!\left(x_{0}x_{0}^{*}\right)\!=\!\!
\sum\nolimits_{\ell=1}^{N}\sum\nolimits_{i=1}^{N}\mathrm{M}_{\ell i}^{(0)}$}. 
Since the linear mapping {\small $\mathrm{vec}^{2}\!\left(\cdot\right)$}
is uniform homeomorphic, \eqref{eq:vec2S_noisy_explicit} implies that
{\small $\mathbf{M}^{(k+1)}$} equals to

\vspace*{-4mm}
%%%%%%%%%%%%%%%%%%%%%%%%%%%%%%%%%%%%%%%%%%%%%%%%%%%%%%%%%%%%%%%%%%%%%%%%%%%%%%%%
\begin{scriptsize}
\begin{equation}\label{eq:S_noisy_explicit}
\mathrm{vec}^{-2}\!\!\left(\!\bm{\Lambda}^{\!k}\mathrm{vec}^{2}\!\left(\!
	\mathbf{M}^{(0)}\!\right)\!\!\right)\!\!+\!\!\!\sum\nolimits_{t=0}^{k-1}\!
	\mathrm{vec}^{-2}\!\!\left(\!{\bm\Lambda}^{\!t}{\bm\Upsilon}{\bm\Xi}^{k-t-1}
	\mathrm{vec}^{2}\!\!\left(\!{\bm \Pi}^{(0)}\!\right)
	\!\!\right)\!.\!\!
\end{equation}
\end{scriptsize}
\vspace*{-6mm}

%%%%%%%%%%%%%%%%%%%%%%%%%%%%%%%%%%%%%%%%%%%%%%%%%%%%%%%%%%%%%%%%%%%%%%%%%%%%%%%%
For {\small $\hat{x}_{0}\!=\!0$}, we have that the first addend in 
\eqref{eq:S_noisy_explicit} produces a null matrix, while the second addend
produces a matrix, denoted by {\small ${\bm W}^{(k)}$}, that is partitioned into 
the blocks of size $n_x$-by-$n_x$, such that 
{\small ${\bm W}^{(k)}\!=\!\Big[W_{\ell i}^{(k)}\Big]_{\ell,i=1}^{N}$}, and

\vspace*{-.5mm}
%%%%%%%%%%%%%%%%%%%%%%%%%%%%%%%%%%%%%%%%%%%%%%%%%%%%%%%%%%%%%%%%%%%%%%%%%%%%%%%%
\begin{scriptsize}
\begin{equation}\label{eq:Qe}
\mathrm{M}_{e}\!=\!\lim\nolimits_{k\to\infty}\sum\nolimits_{\ell=1}^{N}\sum\nolimits_{i=1}^{N}
	W_{\ell i}^{(k)}\!.
\end{equation}
\end{scriptsize}

\vspace*{-6mm}
%%%%%%%%%%%%%%%%%%%%%%%%%%%%%%%%%%%%%%%%%%%%%%%%%%%%%%%%%%%%%%%%%%%%%%%%%%%%%%%%
Therefore, for any initial condition 
{\small $\left(\hat{x}_{0},\hat{\varphi}_{0}\right)$}, i.e., for any 
{\small $\mathbf{M}^{(0)}$}, by the definition of the matrix addition as 
entry-wise sum, it follows that {\small $\lim\nolimits_{k\to\infty}\mathbb{E}(x_{k+1}x_{k+1}^{*})\!-\!M_e\!=\!
0\!=\!\sum\nolimits_{\ell=1}^{N} \sum\nolimits_{i=1}^{N}(\mathrm{M}_{\ell i}^{(k+1)}\!-\!W_{\ell i}^{(k)})$}, 
where {\small $\mathrm{M}_{\ell i}^{(k)}\!\succeq\!0$, $\forall \ell,i,k$}.
This implies that {\small $\forall \mathbf{M}^{(0)}$}, {\small $\lim\nolimits_{k\to\infty}\mathrm{vec}^{-2}(\bm{\Lambda}^{k}\mathrm{vec}^{2}(\mathbf{M}^{(0)}))\!=\!0$},
which holds if and only {\small $\lim\nolimits_{k\to\infty}\!\bm{\Lambda}^{\!k}\!=\!0$},
and thus iff {\small $\rho\!\left({\bm \Lambda}\right)\!<\!1$}. \qed

\subsection{Proof of Theorem \ref{theorem:delayedMeanSquareStabilizability}}
\vspace*{-5mm}
\begin{small}
\begin{align*}
&\text{Define~}\mathbb{V}\!\triangleq\!\Big\{
	{\mathbf{V}_{_{\!1}}}\!=\!\big[V_{{_{\!1}}{\ell i}}\big]_{\ell,i=1}^{N},
	{\mathbf{V}_{_{\!2}}}\!=\!\big[V_{{_{\!2}}{\ell i}}\big]_{\ell,i=1}^{N},
	{\mathbf{V}_{_{\!3}}}\!=\!\big[V_{{_{\!3}}{\ell i}}\big]_{\ell,i=1}^{N}, \\[-1mm]
  &\quad \mathbf{L}\!=\!\big[L_{i}\big]_{i=1}^{N} ~\Big\vert~ 
    V_{{_{\!1}}{\ell i}}\!\in\!\mathbb{F}^{\,n_{x}\times n_{x}}\!, 
	V_{{_{\!2}}{\ell i}}\!\in\!\mathbb{F}^{\,n_{x}\times n_{u}}\!,
	V_{{_{\!3}}{\ell i}}\!\in\!\mathbb{F}^{\,n_{u}\times n_{u}}\!, \\[-1mm]
  &\quad L_{i}\!\in\!\mathbb{F}^{n_{u}\times n_{x}}\!, V_{{_{\!1}}{\ell i}}\!\succ\!0,
	V_{{_{\!3}}{\ell i}}\!\succeq\!0, \text{\;satisfy\;}\eqref{eq:delayedStabilizabilityLMI}~
	\forall \ell,i 
\Big\}.
\end{align*}
\end{small}

\vspace*{-5mm}
%%%%%%%%%%%%%%%%%%%%%%%%%%%%%%%%%%%%%%%%%%%%%%%%%%%%%%%%%%%%%%%%%%%%%%%%%%%%%%%%
Since {\small $V_{{_{\!1}}{ij}}\!\succ\!0$},
\eqref{eq:delayedStabilizabilityLMI_b} is equivalent to 
{\small $L_{i}V_{{_{\!2}}{i j}}\!\succeq\!
V_{{_{\!2}}{ij}}^{*}V_{{_{\!1}}{ij}}^{-1}V_{{_{\!2}}{ij}}$}
(by the Schur complement, see e.g. \cite[Lemma 2.23, p. 28]{costa2006discrete},
\cite[Section A.5.5, pp. 650\,--\,651]{boyd2004convex}).
%%%%%%%%%%%%%%%%%%%%%%%%%%%%%%%%%%%%%%%%%%%%%%%%%%%%%%%%%%%%%%%%%%%%%%%%%%%%%%%%
To prove the necessity, we assume that the system \eqref{eq:mjls} is mean square 
stabilizable. Then, by Definition \ref{def:stabilizability}, there is a mode-dependent 
state-feedback controller {\small $\mathbf{K}\!=\!\left(K_i\right)_{i=1}^{N}$}
such that the system \eqref{eq:mjls_alternative}, with 
{\small $\mathfrak{A}_{\varphi_k}\!\triangleq\!A\!+\!\nu_{\theta_k}BK_{\theta_{k-1}}$}, 
is mean square stable. Then, by Theorem \ref{theorem:delayedMeanSquareStability}, 
{\small $\rho\!\left({\bm \Lambda}\right)\!\!<\!1$}, and by Proposition 
\ref{prop:delayedNoiselessLyapunov},
{\small $\exists \mathbf{Y}\!=\!\!\left[\mathrm{Y}_{ij}\right]_{i,j=1}^{N}$}, 
{\small $\mathrm{Y}_{ij}\!\in\!\mathbb{F}^{n_x\times n_x}$}, 
{\small $\mathrm{Y}_{ij}\!\succ\!0$}, such that
{\small $\mathrm{Y}_{ij}\!-\!\mathcal{L}_{ij}\!\left(\mathbf{Y}\right)
\!\succ\!0, \forall i,j$}, with {\small $\mathcal{L}_{ij}\!\left(\mathbf{Y}\right)$} 
defined by \eqref{eq:operatorLDefinition}. After taking 
{\small $V_{{_{\!1}}{ij}}\!=\!\mathrm{Y}_{ij}$}, 
{\small $V_{{_{\!2}}{ij}}\!=\!V_{{_{\!1}}{ij}}K_{i}^{*}$},
{\small $V_{{_{\!3}}{ij}}\!=\!V_{{_{\!2}}{ij}}^{*}V_{{_{\!1}}{ij}}^{-1}V_{{_{\!2}}{ij}}$}, 
{\small $L_{i}\!=\!K_{i}$},
it is easy to verify from 
{\small $\mathrm{Y}_{ij}\!-\!\mathcal{L}_{ij}\!\left(\mathbf{Y}\right)\!\succ\!0$}
that \eqref{eq:delayedStabilizabilityLMI} are satisfied and therefore the set
{\small $\mathbb{V}$} is not empty.
%%%
%%%%%%%%%%%%%%%%%%%%%%%%%%%%%%%%%%%%%%%%%%%%%%%%%%%%%%%%%%%%%%%%%%%%%%%%%%%%%%%%
To prove the sufficiency, we assume that {\small $\mathbb{V}$} is non empty, 
so there are {\small ${\mathbf{V}_{_{\!1}}}, {\mathbf{V}_{_{\!2}}}$}, 
{\small ${\mathbf{V}_{_{\!3}}}$}, and {\small $\mathbf{L}$}
that satisfy \eqref{eq:delayedStabilizabilityLMI}. 
Let {\small $\mathbf{K}\!=\!\left(K_i\right)_{i=1}^{N}$}, 
{\small $K_i \!\in\!\mathbb{F}^{n_u\times n_x}$}, be such that 
{\small $V_{{_{\!2}}{ij}}\!=\!V_{{_{\!1}}{ij}}K_{i}^{*}$}.
By \eqref{eq:delayedStabilizabilityLMI_b}, such {\small $\mathbf{K}$} exists,
and it may be obtained as {\small $\mathbf{K}\!=\!\mathbf{L}$}. Then, from 
\eqref{eq:operatorLDefinition}, \eqref{eq:delayedStabilizabilityLMI}, we have that 
{\small $\mathcal{L}_{ij}\!\left({\mathbf{V}_{_{\!1}}}\!\right)\!-\!V_{{_{\!1}}{i j}}$}
equals to 

\vspace*{-4mm}
%%%%%%%%%%%%%%%%%%%%%%%%%%%%%%%%%%%%%%%%%%%%%%%%%%%%%%%%%%%%%%%%%%%%%%%%%%%%%%%%
\begin{scriptsize}
\begin{align*}
&\!\Bigg(\!\!A\!\!\sum\nolimits_{\ell=1}^{N}\!\!\!\!V_{{_{\!1}}{\!\ell i}}A^{*}\!\!\!+\!\!B\!\!
\sum\nolimits_{\ell=1}^{N}\!\!\!\!\!K_{\ell}\!V_{{_{\!1}}{\!\ell i}}K_{\ell}^{*}\!B^{*}\!\hat{\nu}_i\!\!+\!\!
2\Re\!\Bigg(\!\!A\!\!\sum\nolimits_{\ell=1}^{N}\!\!\!V_{{_{\!1}}{\!\ell i}}K_{\ell}^{*}\!B^{*}\!
\hat{\nu}_i\!\!\Bigg)\!\!\!\Bigg)p_{ij}\!\!-\!\!V_{{_{\!1}}{i j}}\!=\\[-1mm]
&\sum\nolimits_{\ell=1}^{N}p_{ij}\!\left(AV_{{_{\!1}}{\!\ell i}}A^{*}\!\!\!+\!\!
BV_{{_{\!3}}{\ell i}}B^{*}\!\hat{\nu}_i\!\!+\!\!
AV_{{_{\!2}}{\ell i}}B^{*}\!\hat{\nu}_i\!\!+\!\!
BV_{{_{\!2}}{\ell i}}^{*}A^{*}\!\hat{\nu}_i\right)\!-\!V_{{_{\!1}}{i j}}\!\prec\!0.
\end{align*}
\end{scriptsize}

\vspace*{-4mm}
So, by Proposition \ref{prop:delayedNoiselessLyapunov}, we have that 
{\small $\rho\!\left({\bm \Lambda}\right)\!<\!1$}, and thus, by 
Theorem \ref{theorem:delayedMeanSquareStability}, the system \eqref{eq:mjls_alternative}
is mean square stable. Then, by Definition \ref{def:stabilizability}, the system
\eqref{eq:mjls} is mean square stabilizable, and the proof is concluded. \qed
 
%%%%%%%%%%%%%%%%%%%%%%%%%%%%%%%%%%%%%%%%%%%%%%%%%%%%%%%%%%%%%%%%%%%%%%%%%%%%%%%%
\subsection{Proof of Theorem \ref{theorem:solModifiedARE}}
By hypothesis, {\small $\left(\mathbf{p}_i\right)_{i=1}^{N}$} is the stationary distribution
of the channel states, and {\small $\hat{\nu}\!=\!\sum_{j=1}^{N}\mathbf{p}_j\hat{\nu}_j$}.
By definition of the steady state distribution, 
{\small $\sum_{i=1}^{N}\mathbf{p}_i\!=\!1$}, and 
{\small $\mathbf{p}_j\!=\!\sum_{i=1}^{N} \mathbf{p}_i p_{ij}$}, so 
{\small $\hat{\nu}\!=\!\sum_{i=1}^{N}\mathbf{p}_{i}\sum_{j=1}^{N}p_{ij}\hat{\nu}_j$}.
Thus, the MARE can be written as {\small $\sum_{i=1}^N\mathbf{p}_{i}\Big(X_{\infty}^b-\!A^{*}X_{\infty}^b A - Q \,+ \sum_{j=1}^{N}p_{ij}\hat{\nu}_j\big(A^{*} X_{\infty}^b B\big)\!\big(R\!+\!B^{*} 
	X_{\infty}^b B\big)^{{-1}}\!\big(B^{*} X_{\infty}^b A\big)\!\Big)\!=\!0$},
holding
{\small $\forall \left(\mathbf{p}_i\right)_{i=1}^{N}$} iff, {\small $\forall i\!\leq\!N$} the following
expression is satisfied: 

\begin{small}
\begin{equation}\label{eq:proofTh1}
X_{\infty}^b = A^{*}X_{\infty}^b A + Q - \!\left(\sum\nolimits_{j=1}^{N}p_{ij}\hat{\nu}_j\!\right) \!Y,
\end{equation}
\end{small}

\noindent
where {\small $Y\!\triangleq\!\big(A^{*} X_{\infty}^b B\big)\!\big(R\!+\!B^{*} 
	X_{\infty}^b B\big)^{{-1}}\!\big(B^{*} X_{\infty}^b A\big)\!$}.
Since {\small $p_{ij}$} and {\small $\hat{\nu}_j$} are known scalars,
{\small $\sum\nolimits_{j=1}^{N}p_{ij}\hat{\nu}_j\!\triangleq\!\xi_i$}, with
{\small $\xi_i$} again a known scalar. So, we focus on the 
term {\small $\xi_i Y$}. From the property of the product of invertible matrix with a non-zero scalar, 
it follows that {\small $\xi_i Y \!=\! \big(\xi_i A^{*} X_{\infty}^b B\big)\!\big(\xi_i(R\!+\!B^{*} 
	X_{\infty}^b B)\big)^{{-1}}\!\big(\xi_i B^{*} X_{\infty}^b A\big)$}.
%\begin{small}
%\begin{equation*}
%\xi_i Y \!=\! \big(\xi_i A^{*} X_{\infty}^b B\big)\!\big(\xi_i(R\!+\!B^{*} 
%	X_{\infty}^b B)\big)^{{-1}}\!\big(\xi_i B^{*} X_{\infty}^b A\big).
%\end{equation*}
%\end{small}
Thus, we apply the definition of {\small $\xi_i$} and substitute the last expression of 
{\small $\xi_i Y$} in \eqref{eq:proofTh1}, obtaining exactly \eqref{eq:care}, where,
as required by the mode-independence,  
{\small $X_{\infty,i}^{c}\!=\!X_{\infty}^b$}, $\forall i\!\leq\!N$.
% \!=\!\hat{X}_{\infty}^c
{\small \qed}
\end{document}